\documentclass[aps,pre,reprint,groupedaddress]{revtex4-2}

\usepackage{amsmath}
\usepackage{amssymb}
\usepackage{graphicx}
 \usepackage{epstopdf}
\usepackage{float}

\usepackage{placeins}
\usepackage{mathtools}
\usepackage{bm}
\usepackage{array,multirow}

\usepackage{xcolor}
\usepackage{booktabs}

\usepackage{trfsigns}

\usepackage[normalem]{ulem}

\begin{document}


\title{Extreme value theory for constrained physical systems}


\author{Marc H\"oll, Wanli Wang and Eli Barkai}
\affiliation{Department of Physics, Institute of Nanotechnology and Advanced Materials, Bar-Ilan University, Ramat-Gan 52900, Israel}



\begin{abstract}
We investigate extreme value theory for physical systems with a global conservation law which describe renewal processes, mass transport models and long-range interacting spin models. As shown previously, a special feature is that the distribution of the extreme value exhibits a non-analytical point in the middle of the support. We expose exact relationships between constrained extreme value theory and well-known quantities of the underlying stochastic dynamics, all valid beyond the midpoint in generality, i.e. even far from the thermodynamic limit. For example for renewal processes, the distribution of the maximum time between two renewal events is exactly related to the mean number of these events. In the thermodynamic limit, we show how our theory is suitable to describe typical and rare events which deviate from classical extreme value theory. For example for the renewal process, we unravel dual scaling of the extreme value distribution, pointing out two types of limiting laws: a normalisable scaling function for the typical statistics and a non-normalised state describing the rare events.
  \end{abstract}


\maketitle


\section{Introduction}

Extreme events are a large class of phenomena in natural and man-made systems which are uncommon compared to the usual dynamics \cite{bouchaud1997universality,albeverio2006extreme,embrechts2013modelling,fortin2015applications,fortin2015applications,MAJUMDAR20201}. Despite their rare occurrence they still can have influential consequences, e.g. the fastest sperm in fertilization \cite{meerson2015mortality,schuss2019redundancy}, the longest trapping time in transport \cite{wang2019transport} and first passage problems in Markov processes \cite{hartich2019extreme}. The original problem considers a set of $N\in\mathbb{N}$ independent and identically distributed (IID) random variables $(x_1,\ldots,x_N)$ and describes the statistics of its maximum $x_\text{max}=\text{max}(x_1,\ldots,x_N)$. Let $\psi(x)$ be the probability density function (PDF) of the random variables and $\Psi(x)$ the cumulative distribution function (CDF). When the maximum $x_\text{max}$ has the value $m$ then all other random variables are less than or equal to $m$. So the CDF of the maximum is $\text{Prob}(x_\text{max}\le m)=\Psi^N(m)$ and hence the PDF of the maximum is obviously
\begin{equation}\label{evt}
f(m) = N \psi(m) \Psi^{N-1}(m).
\end{equation}
A central result of classical extreme value theory (EVT) is that the limiting maximum PDF for large $N$ converges to one of three classes of distributions called Weibull, Gumbel or Fr\'echet depending on the large $x$ behaviour of $\psi(x)$ when $m$ is shifted and rescaled appropriately \cite{embrechts2013modelling,zarfaty2020accurately,
gumbel2012statistics,fisher1928limiting}. However, for most systems the assumption of IID random variables has to be abandoned.\\

Recently EVT was studied for a wide range of different models whose common property is the global confinement of their dynamics, see \cite{MAJUMDAR20201} for a review. This global conservation induces correlations among the random variables. It is a common trait shared in many models including \textit{renewal processes} (\textit{RP}) \cite{godreche2001statistics,niemann2016renewal,wang2018renewal,feller1971introduction,lowen1993fractal}, mass transport models such as \textit{zero range processes} (\textit{ZRP}) \cite{evans2006canonical,majumdar2010real,majumdar2005nature,zia2004construction,evans2005nonequilibrium}, and long-range interacting spin models such as the \textit{truncated inverse distance squared Ising model} (\textit{TIDSI}) \cite{bar2014mixed,bar2014mixed2}. These three models describe numerous physical systems, including zero crossing of Brownian motion, arrival times at a detector, interacting systems to name only a few. \\

Particular attention was devoted to systems which loosely speaking are scale free, such as fractal renewal theory with diverging mean waiting time and diverging variance of the waiting time (see below). These systems exhibit large fluctuations and dominance of the extreme. It was shown previously how the global constraint may modify completely the classical EVT in the sense of strong deviations from Fr\'echet's law. Somewhat similar to the classical ensembles of statistical physics, e.g. microcanonical ensembles with fixed
energy, volume and number of particles and canonical ensembles where the temperature of the bath is the constrain, the different constraints discussed below also give rich physical behaviours specific to the ensemble/model. For each model there are several classes of limiting laws in the thermodynamic limit when the global constraint diverges. These classes depend on the model parameters and were studied for \textit{RP} \cite{godreche2009longest,godreche2014universal,godreche2015statistics,
vezzani2019single,wang2019transport,scheffer1995rank}, \textit{ZRP} \cite{majumdar2010real,evans2008condensation} and \textit{TIDSI} \cite{bar2016exact}. For example for \textit{RP} with fat-tailed waiting times, typical fluctuations of the maximum go through a dynamical phase transition depending on the existence or non-existence of the mean waiting time. When the mean exists Fr\'echet's law holds typically, when it doesn't exist the behaviour is completely different \cite{godreche2015statistics}. A similar situation exists for \textit{TIDSI} in the critical phase between ferromagnetic and paramagnetic phase \cite{bar2016exact,godreche2017longestaa}. However, a particular limiting law might reflect only part of the truth. So does Fr\'echet's law predict a diverging second moment of the largest waiting time in a renewal process. However, that is impossible since all waiting times are shorter than the observation time. This does not imply that Fr\'echet's law is incorrect, only that it must be modified in its tail. To put differently, the constraint yields a natural cut off  and this modifies the description of classical EVT \cite{vezzani2019single,wang2019transport}. In below main text, we discuss the thermodynamic limit for each model and how our results help to classify limiting behaviours. \\

Our work addresses two main themes. First without restoring to a thermodynamic limit, we provide complete set of relations between constrained EVT and much simpler quantifiers of the underlying stochastic dynamics. These exact relations are valid for any value of the global constraint, i.e. close to and far from the thermodynamic limit. For example, for \textit{RP} we find an exact and simple relation between EVT and the mean number of renewals $\langle N \rangle$. The relations are found beyond the critical point $m>C/2$ where $C>0$ is the global constraint. It has been recognized in earlier studies \cite{wendel1964zero,godreche2017longestaa} that the analysis beyond this midpoint may be simplified for the Brownian bridge and the tied-down renewal process (which is essentially \textit{TIDSI}). The importance of the midpoint is easy to understand: Once we observe a value larger than half of the global constraint it is already the maximum. No following value can be larger. \\

Our second goal is to exploit the exact relations and consider the thermodynamic limit. We recap known and also find new limiting laws. For example for renewal processes, we find dual scaling, i.e. our theory describes both types of limiting behaviour. When no moment of the waiting times exist, our theory describes typical events and rare events. When only the first moment of the waiting times exist, our theory describes the correction to Fr\'echet's law (considered as rare events) and its large deviations. In this sense we go beyond previous studies of the thermodynamic limit \cite{MAJUMDAR20201,evans2008condensation,godreche2015statistics,bar2016exact,
vezzani2019single,wang2019transport}. We further confirm that rare events can be often described within the framework of infinite densities \cite{vezzani2019single,
wang2019transport,
rebenshtok2014non,
akimoto2020infinite,
kessler2010infinite,
wang2019ergodic,
akimoto2015distributional,
akimoto2010role}.\\ 

The article is constructed as follows. We consider the \textit{RP} in section \ref{sec:rp}, the \textit{ZRP} in \ref{sec:zrp} and the \textit{TIDSI} in section \ref{sec:tidsi}. For all three models we derive the maximum PDF in the second half of the support and relate it to well-studied stochastic quantifiers of the underlying dynamics. There we present the analysis on the \textit{RP} elaborately. Furthermore, for the \textit{RP} and \textit{TIDSI} we derive limiting laws of the second half maximum distribution in the thermodynamic limit for fat-tailed random variables. Section \ref{sec:conclusion} gives a summary.

\section{Renewal process}\label{sec:rp}

\subsection{Basics}

\textit{RP} are widely used in physics \cite{godreche2001statistics,
godreche2015statistics,
niemann2016renewal,
wang2018renewal,
feller1971introduction,
lowen1993fractal}, for example in describing the random arrival times of radioactive debris to a Geiger counter. Mathematically these processes are described with a PDF $\psi(\tau)$  of inter-arrival times, sometimes called waiting times. The process starts at time $t_1=0$ considered as the first event. To construct the process, first, sample $\tau_1$ from the PDF $\psi(\tau)$ (this describes the timing of the second event), then renew the process by sampling $\tau_2$ from the same PDF so that the timing of the third event is given by $\tau_1+\tau_2$. The process is continued this way for $N$ events, i.e. the $i$-th event happens at time $t_i=\tau_1+\tau_2+\ldots \tau_{i-1}$ with $i\in \{1,2,\ldots\}$. The waiting times $\tau_1,\tau_2,\ldots$ between events are IID random variables all sampled from $\psi(\tau)$. The PDF of $\psi(\tau)$ can be either thin-tailed or fat-tailed and this has major consequences on the behaviour of the extreme events. For example, an exponential (thin-tailed) PDF $\psi(\tau)$ describes arrival times of independent photons to a detector. An example of a fat-tailed process is the zero crossing of Brownian motion where $\psi(\tau)\sim \tau^{-3/2}$, similarly for blinking quantum dots \cite{stefani2009beyond,margolin2005nonergodicity} or times between jumps in the anomalous continuous time random walks \cite{metzler2000random,kutner2017continuous}. \\

 The renewal process is observed at the observation time $t=T$. The fixed observation time $T$ is the sum of all waiting times before the last event $i=N$ added with the backward recurrence time 
\begin{equation}\label{fixedmeasurementtime}
T=\sum_{i=1}^{N-1} \tau_i + \tau_B,
\end{equation}
see Fig.~(\ref{fig:modelconc}). The backward recurrence time $\tau_B$ is the time interval between the last event $i=N$ and the observation time $T$. It is differently distributed than the waiting times \cite{cox1962renewal}. The constraint of a fixed observation time implies that the amount of events $N$ is a random number. This and the cut off of the last time interval $\tau_{N}$ to $\tau_\text{B}$ make the set of all waiting times $\{\tau_1,\tau_2,\ldots,\tau_{N-1},\tau_\text{B}\}$ non-IID.

\subsection{Overview of constrained models}
\begin{center}
\begin{figure}[ht]
\includegraphics[width=0.45\textwidth]{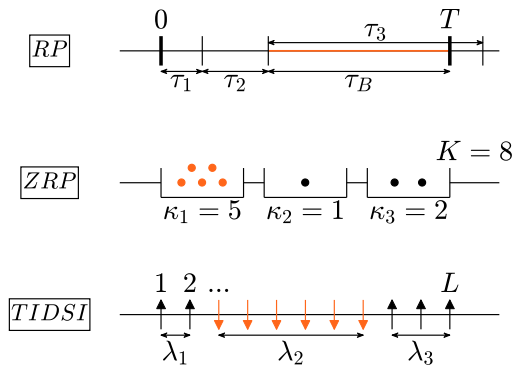}
\caption{Schematic figure of the three models \textit{renewal process}, \textit{zero range process} and \textit{truncated inverse distance squared Ising model} presented in the main text. The maximum in each model is colored organge. The \textit{ZRP} is described in section \ref{sec:tidsi} and \textit{TIDSI} in section \ref{sec:tidsi}.\label{fig:modelconc}}
\end{figure}
\end{center}

Before we continue with the maximum statistics of the waiting times $\{\tau_1,\tau_2,\ldots,\tau_{N-1},\tau_\text{B}\}$, we compare model details of \textit{RP} with the two later studies models, \textit{ZRP} and \textit{TIDSI}. The common trait of these models is that the sum of the random variables
\begin{equation}
C=\sum_{i=1}^N x_i 
\end{equation}
is fixed to the global constraint $C>0$. For example for the \textit{RP}, the constraint is the observation time $C=T$ and the random variables are $x_i=\tau_i$ when $i\in[1,N-1]$ and $x_N=\tau_B$. In Table \ref{tab:table01} important characteristics are presented with the appropriate $x_i$ and $C$ for each model. Fig.~\ref{fig:modelconc} shows a schematic figure of the three models. We do not define precisely \textit{ZRP} and \textit{TIDSI} at this stage, we will do so later in section \ref{sec:zrp} and \ref{sec:tidsi}. For now we just want to define their global constraint: For \textit{ZRP}, $C$ is the total number of particles in a systems where particles are distributed in boxes, while for \textit{TIDSI} describing an interacting spin system $C$ is the size of the system. In both models one can say that interactions are local, i.e. only particles within a box interact and only spins within a given domain.

\begin{table}[H]\centering
  \begin{tabular}{ l || l | l | l }

& \textit{RP} & \textit{ZRP} & \textit{TIDSI}\\ \hline
Random & Waiting & Number of & Domain \\
variables $x_i$ & times $\tau_i$ & particles $\kappa_i$ & lengths $\lambda_i$ \\ \hline
Values of $x_i$ & Continuous & Discrete & Discrete \\ \hline
Constraint $C$ & Observation time  & Total number & Total length\\ 
& $T=\sum\limits_{i=1}^{N-1}\tau_i+\tau_B$ & $K=\sum\limits_{i=1}^N\kappa_i$ & $L=\sum\limits_{i=1}^N \lambda_i$\\ \hline
Number $N$ & Random & Fixed & Random 

  \end{tabular}\caption{\label{tab:table01}Overview of details of the three models \textit{renewal process}, \textit{zero range process} and \textit{truncated inverse distance squared Ising model} presented in the main text. Their relevant random variables, the constraint and the randomness of $N$ are shown. The \textit{ZRP} is described in section \ref{sec:tidsi} and \textit{TIDSI} in section \ref{sec:tidsi}.}\end{table}

\subsection{Extreme value statistics}

We investigate the statistics of the maximum waiting time \cite{godreche2015statistics,vezzani2019single,wang2019transport}
\begin{equation}
\tau_\text{max} = \text{max}(\tau_1,\tau_2,\ldots,\tau_{N-1},\tau_\text{B}).
\end{equation}
The maximum $\tau_\text{max}$ is also called extreme event of the waiting times. The maximum PDF is defined by $f(m;T) = dF(m;T)/dm$. The maximum CDF $F(m;T)= \text{Prob}(\tau_\text{max} \le m)$ is the probability of the random variable $\tau_\text{max}$ being less than or equal $m$. Clearly, the maximum is constrained $0<m\le T$. Since the number of events $N$ is random it is instructive to consider
\begin{equation}\label{sumcum}
f(m;T) = \sum_{N=1}^\infty f_N(m;T)
\end{equation}
with $f_N(m;T)=dF_N(m;T)/dm$ being the maximum PDF with exactly $N$ renewal events. In this context the value of $N$ is a sampled value. The maximum CDF with exactly $N$ events is given by \cite{godreche2015statistics}
\begin{equation}\begin{split}\label{singlecum}
F_N(m;T)&=\int\limits_0^m\ldots  \int\limits_0^m  \int\limits_0^m \psi(\tau_1)\ldots\psi(\tau_{N-1})\varphi(\tau_B)\\
&\times \delta \left(T-\left(\sum_{i=1}^{N-1}\tau_i+\tau_B\right)\right) d\tau_1 \ldots d\tau_{N-1} d\tau_B.
\end{split}\end{equation}
This is the probability of $\tau_\text{max}$ being less than or equal to $m$ when exactly $N$ events happened. Here, the survival probability
\begin{equation}
\varphi(\tau_B)=\int\limits_{\tau_B}^\infty \psi(\tau)d\tau
\end{equation}
is the probability that no other event than the first one at $t_1=0$ occurs until time $\tau_B$. Eq.~(\ref{singlecum}) is easy to interpret, the set of waiting times $\{\tau_1,\ldots,\tau_B\}$ are all less than or equal to $m$, and the delta function is the constraint. Since we will use this formula below we write the $N$- multiple integral shorter as 
\begin{equation}\label{singlecum2}
F_N(m;T)=\int\limits_0^m \psi(\tau_1)\ldots\psi(\tau_{N-1}) \varphi(\tau_N)\delta(T-\|\bm{\tau}\|_1)  d\bm{\tau}
\end{equation}
with the $N$-vector $\bm{\tau}=(\tau_1,\ldots,\tau_N)^\text{T}$ and the taxicab norm $\|\bm{\tau}\|_1=\sum_{i=1}^N\tau_i$.  \\ \par

\begin{center}
\begin{figure*}
\includegraphics[width=0.9\textwidth]{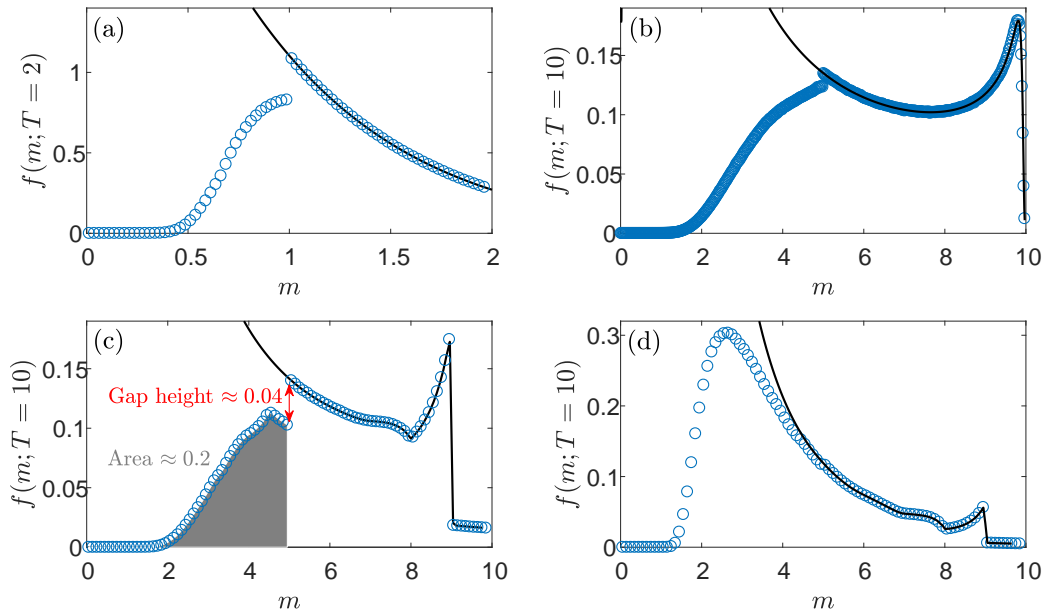}
  \caption{Histogram of the maximum PDF $f(m;T)$ of \textit{RP} from Monte Carlo simulations (blue circles) compared with the theory of Eq.~(\ref{formal0}) (black line) for (a) exponential $\psi(\tau)=\text{exp}(-\tau)$ with $T=2$, (b) one-sided L\'evy $\psi(\tau)=1/(\sqrt{2\pi})\tau^{-3/2}\text{exp}(-1/(4\tau)$ with $T=10$ and (c),(d) Pareto $\psi(\tau)=\alpha \tau^{-\alpha}$, $\tau>1$, with $T=10$. The simulations were performed with $10^7$ realizations. The analytical expressions of $\langle N \rangle$ and $R$ were obtained via numerical inverse Laplace transform, see Eq.~(\ref{largetau}). The non-analytical point $T/2$ is well visible in (a), (b) and (c). It is barely visible for $\alpha=3/2$ in (d) because $T$ is relatively large. In addition, we show the theoretical gap height Eq.~(\ref{gapheight}) and the theoretical area of the first half Eq.~(\ref{areafirst}) exemplary in (c), both match with their numerical estimates.}\label{fig:four}
\end{figure*}
\end{center}

Before we continue our analytical investigation, let us take a look at simulation results with which we construct the PDF of $\tau_\text{max}$. In Fig.~\ref{fig:four} we simulate the process and obtain the histograms for $f(m;T)$ where we used the exponential waiting time PDF $\psi(\tau)=\text{exp}(-\tau)$, the Pareto waiting time PDF $\psi(\tau)=\alpha\tau^{-1-\alpha}$ with $\tau \ge 1$ and the one-sided L\'evy waiting time PDF $\psi(\tau)=1/(2\sqrt{\pi})\tau^{-3/2}\text{exp}(-1/(4\tau))$. All examples show a discontinuity at the midpoint of the support
\begin{equation}
m=\frac{T}{2}.
\end{equation}
The Pareto waiting time PDF also yields in an intrinsic discontinuity at $m=9$ because $\tau\ge 1$. The importance of $T/2$ can be intuitively understand: Once a waiting time is larger than $T/2$ it is then also the maximum waiting time. No previous and following waiting time can be larger. Since the PDF of $\tau_\text{max}$ is non-analytical we cannot expect to find a global solution in the whole range $0<m<T$. Importantly, all the results in Fig.~\ref{fig:four} are presented for finite simulation time and far from the thermodynamic limit. E.g. for the exponential process the mean waiting time is unity and the measurement time is just twice as large. Usually one does not expect general statistical laws to emerge at this limit. However, in the Fig.~\ref{fig:four} we show a theory (derived  below) that works perfectly beyond the mid point. Our goal is then to present this theory and only later consider the thermodynamic limit.

\subsection{Extreme value statistics in the second half}

We now present the main result and its derivation afterwards. In the second half $T/2<m<T$, the maximum PDF $f(m;T)$ is exactly related to the mean number of renewals $\langle N(t) \rangle$, namely
\begin{equation}\label{rppdfnew}
\boxed{f(m;T) = \varphi(m) R(T-m)+\psi(m) \langle N(T-m) \rangle.}
\end{equation}
The function $R(T-m)$ is the rate of producing these events, namely the derivative of $\langle N \rangle$. Both $\langle N \rangle $ and $R$ are thoroughly investigated in the physical and mathematical literature \cite{godreche2001statistics,niemann2016renewal,wang2018renewal,feller1971introduction}. We find also an elegant formula of the maximum CDF
\begin{equation}\label{rpcdfnew}
\boxed{F(m;T)=1- \varphi(m) \langle N(T-m) \rangle}
\end{equation}
again in the second half $T/2<m<T$. The formulas Eq.~(\ref{rppdfnew}) and (\ref{rpcdfnew}) present exact results for any $m>T/2$ and are very useful as it allows us to derive both finite time expressions and also the long-time limit (see below).\\

 Both relationships yield insight on the maximum $m$ when it is roughly of order $T$. Then we need to have information on $\langle N(T-m) \rangle$, which includes $R(T-m)$, only for very short time. Namely, once we have $\langle N(T-m) \rangle$ for time $T-m$ we can predict EVT for $m$. Intuitively, to observe a large $m$ of order $T$, we need the maximum to be produced close to the start of the process.\\

We will now present the derivation of this main result. Taking the derivative of $F_N(m;T)$ from Eq.~(\ref{singlecum}) or (\ref{singlecum2}) yields two terms with each a $(N-1)$-multiple integral 
\begin{equation}\begin{split}\label{fint}
&f_N(m;T)=\underbrace{\varphi(m)\int\limits_0^m \prod_{i=1}^{N-1}\psi(\tau_i) \delta(T-m-\|\bm{\tau}\|_1)  d\bm{\tau}}_{\mathcal{B}}\\
&+\underbrace{\psi(m)(N-1)\int\limits_0^m \prod_{i=1}^{N-2}\psi(\tau_i)\varphi(\tau_{N-1})\delta(T-m-\|\bm{\tau}\|_1)  d\bm{\tau}}_{\mathcal{NB}}.
\end{split}\end{equation}
Here $\bm{\tau}=(\tau_1,\ldots,\tau_{N-1})^\text{T}$. The first term describes backward and the second term non-backward processes:
\begin{equation}\begin{split}
\mathcal{B} &= \text{Renewal processes with } \tau_\text{max} = \tau_B,\\
\mathcal{NB} &= \text{Renewal processes with } \tau_\text{max} \neq \tau_B.
\end{split}\end{equation}
The two integrals in Eq.~(\ref{fint}) are special cases of this general integral
\begin{equation}\label{inttest}
I_{N-1}(m,T^\prime)=\int\limits_0^m \prod_{i=1}^{N-1} g_i(\tau_i) \delta\left(T^\prime - \|\bm{\tau} \|_1 \right) d\bm{\tau}.
\end{equation}
We assume general positive functions $g_i$ and an arbitrary constant $T^\prime>0$. Compared with Eq.~(\ref{fint}) it is $T^\prime=T-m$ and the functions $g_i$ are either $\psi$ or $\varphi$. When we restrict the regime $T^\prime<m$ (for Eq.~(\ref{fint}) it means $T/2<m$) then this integral is identical to the $(N-1)$- fold convolution
\begin{equation}\label{convotrick}
I_{N-1}(m,T^\prime) = (g_1\ast\ldots \ast g_{N-1})^{(N-1)}(T^\prime)
\end{equation}
which we proof rigorously in Appendix \ref{sec:appa}. The $2$-fold convolution is defined as $(g_1\ast g_2)(T^\prime)=\int_0^{T\prime} g_1(\tau_1)g_2(T^\prime-\tau_1)d \tau_1$ and higher order convolutions are obtained successively. Eq.~(\ref{convotrick}) means that the upper limit of the integration $m$ is reduced to $T^\prime<m$. That is because the constraint $T=\|\bm{\tau} \|_1$ forces all individual $\tau_i$ to be less than $T^\prime$. Therefore the integration from $T^\prime$ to $m$ yields zero. Hence we remain with the convolution. In Appendix \ref{sec:appa} we show Eq.~(\ref{convotrick}) in detail. Importantly, we realise that this decoupling trick, valid whenever $T^\prime<m$, is a very general theme. We use this trick also below for the two other models, i.e the \textit{ZRP} and the \textit{TIDSI}.\\

We see now why we consider the maximum PDF instead of the maximum CDF. The delta function of the maximum PDF depends on $T-m$. We set now $T^\prime = T - m$ and apply Eq.~(\ref{convotrick}) onto Eq.~(\ref{fint}) under the assumption of the second half $T/2< m < T$, we obtain exactly
\begin{equation}\begin{split} \label{fn}
&f_N(m;T) \\
&= \varphi(m) Q_{N-1}(T-m) + \psi(m) (N-1) P_{N-1}(T-m).
\end{split}\end{equation}
Here we introduced two quantities well-known from renewal theory \cite{akimoto2020infinite}. The first quantity is the distribution $Q_{N-1}(t)=\left\langle \delta\left(t-\|\bm{\tau}\|_1 \right) \right\rangle$ of having the $N$-th renewal event exactly at time $t$. It can be written as the iteration equation
\begin{equation}
Q_{N-1}(t) = (Q_{N-2}\ast\psi)(t)
\end{equation}
with $Q_0(t)=\delta(t)$. The second quantity $P_{N-1}(t)$ is the probability of finding $N-1$ renewal events up to time $t$. Both are connected with the survival probability via
\begin{equation}\label{ewq}
P_{N-1}(t)=(Q_{N-1}\ast\varphi)(t)
\end{equation}
In case of a single event process $N=1$ we get from Eq.~(\ref{fn}) that 
\begin{equation}\label{zerodelta}
f_1(m;T)=\varphi(m)\delta(T-m)
\end{equation}
which describes the delta peak of the maximum PDF $f(m;T)$ at $m=T$.\\

The first term of Eq.~(\ref{fn}) means that the last waiting time $\tau_B$ is maximum and the second term describes the $N-1$ other cases where the maximum ended before $T$. When the last waiting time is maximum then at time $T-m$ exactly $N$ events happened which gives $Q_{N-1}(T-m)$. This is multiplied with the probability of not having an event during $m$, namely $\varphi(M)$. Hence we have the first term. Now the second term consists of $\psi(m)$, i.e. the maximum ended before $T$, and $(N-1)P_{N-1}(T-m)$. It simply means that we had $N-1$ events in the remaining time $T-m$. Note that any of the $N-1$ waiting times excluding the backward recurrence time might be the largest, so the second term is multiplied by $N-1$. Hence we have the second term.\\ 

Summing up all number of events in Eq.~(\ref{fn}) yields the main result of this section, namely the maximum PDF when $T/2< m< T$ exactly given by
\begin{equation}\label{formal0}
f(m;T) = \varphi(m) R(T-m) + \psi(m) \langle N(T-m)\rangle.
\end{equation}
Thus, we derived Eq.~(\ref{rppdfnew}). The first term contains the rate function
\begin{equation}\label{fout}
R(T-m)=\sum_{N=1}^\infty Q_{N-1}(T-m)
\end{equation}
which is the probability of finding some event exactly at time $T-M$, \cite{akimoto2020infinite}. The delta function $\delta(T-M)$ from $N=1$ does not contribute since $m<T$, and further we have already pointed out the
behaviour of solution when the maximum is equal the observation time Eq.~(\ref{zerodelta}). The second term in Eq.~(\ref{formal0}) contains the mean number of renewal events
\begin{equation}
\langle N(T-m)\rangle =\sum_{N=1}^\infty N P_N(T-M)
\end{equation}
It is related to the rate function via the definite integral
\begin{equation}\label{defint}
\langle N(T-M)\rangle = \int_0^{T-M} R(t)dt.
\end{equation}
Note that if $T$ is large but we limit ourselves to rare events when also $m$ is large,
such that $T-m$ is small, Eq. (\ref{formal0}) states that all we need to evaluate is the short-time behaviour $R$ and $\langle N\rangle$. \\

Although $R$ and $\langle N\rangle$ are well investigated observables within renewal theory, still an exact and explicit analysis of Eq.~(\ref{formal0}) is difficult due to the convolutions. As well-known, it is beneficial to analyse such problems in Laplace space. The Laplace transform of some function $h(t)$ is defined by
\begin{equation}
\widehat{h}(s) = \mathcal{L}_{t\to s}\{h(t)\}=\int\limits_0^\infty h(t) e^{-st}dt.
\end{equation}
The Laplace transform of Eq.~(\ref{formal0}) with respect to the observation time $T$ is
\begin{equation}\label{largetau}
\widehat{f}(m;s) = \varphi(m) \frac{e^{-sm}}{1-\widehat{\psi}(s)} + \psi(m)\frac{e^{-sm}}{s(1-\widehat{\psi}(s))}
\end{equation}
which is easy to prove with the convolution theorem of Laplace transforms. In detail, we used $\widehat{Q}_N(s) = \widehat{\psi}^N(s)$ and $\widehat{P}_N(s)=\widehat{\psi}^{N-1}(s)(1-\widehat{\psi}(s))/s$ and the geometric series. Note that Eq.~(\ref{largetau}) is only valid for inverse Laplace transforms $\mathcal{L}^{-1}_{s\to T}$ when $T/2 < m<T$.

\subsection{Maximum CDF in the second half}

\begin{center}
\begin{figure}[ht]
\includegraphics[width=0.44\textwidth]{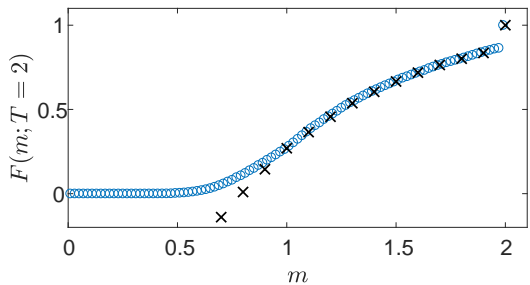}
\caption{Comparison of two approaches to estimate the CDF $F(m;T)$ with $T=2$ for exponential $\psi(\tau)=\text{exp}(-\tau)$: first by direct estimation (blue circles) and secondly by estimation of $\langle N \rangle$ and using Eq.~(\ref{rpcdf}) (black crosses). The number of simulations is $10^4$. As predicted by Eq.~(\ref{rpcdf}), both curves match for the second half $T/2<m<T$. At $m=T$ the discontinuity is also captured by both curves because $\langle N(0) \rangle=0$. Eq.~(\ref{rpcdf}) can also be calculated analytically as $1-\text{exp}(T-m+1)$.\label{fig:rpcdf}}
\end{figure}
\end{center}

From normalization, the maximum CDF at $m=T$ is clearly $F(T;T)=1$. The CDF is discontinuous due to samples with the only renewal at $t_1=0$.  We separate the contribution from these realisations with single renewal events, described by Eq.~(\ref{zerodelta}), and the remaining processes where we had at least two renewals. Therefore we have 
\begin{equation}
 \lim_{m\to T} F(m;T)+\varphi(T)=1
\end{equation}
where $F(T;T)=\varphi(T)$ is the probability of $m=T$. With this boundary condition, we may integrate the PDF $f(m;T)$ and then get the maximum CDF for $T/2 < m< T$ as
\begin{equation}\label{rpcdf}
F(m;T) = 1 - \varphi(m) \langle N(T-m) \rangle.
\end{equation}
Thus, we derived Eq.~(\ref{rpcdfnew}). In Fig.~\ref{fig:rpcdf} we simulate both sides of this formula and find perfect matching the second half. Especially the estimation of the right hand side, i.e. estimation of $\langle N \rangle$ and putting into $1 - \varphi(m) \langle N(T-m) \rangle$, demonstrates 

In principle, the same can be done for $f$. \\

Specifically the probability of finding the maximum $\tau_\text{max}$ in the first half time $m<T/2$ is
\begin{equation}\label{areafirst}
F(T/2;T)=1-\varphi(T/2)\langle N(T/2) \rangle 
\end{equation}
which is valid for all waiting time PDFs $\psi(\tau)$.

\subsection{Gap height of the maximum PDF at the mid-point}

The non-analytical behavior at $T/2$ arises from double event processes with $N=2$ renewals. The set of waiting times is $(\tau_1,\tau_B)$. The maximum of this set is always larger than $T/2$, i.e. the probability of $m<T/2$ is zero. We can quantify this with
  \begin{equation}\begin{split}
  f_2(M;T)
  =
\begin{cases}
0 & \text{ if }m<T/2,\\
   \varphi(m) \psi(T-m) \\
   + \psi(m) \phi(T-m) & \text{ if } m> T/2
\end{cases}  
 \end{split}\end{equation}
 which is derived from Eq.~(\ref{fint}). Thus the height of the gap between the first and second half time expression of $f(M;T)$ at $T/2$ is
\begin{equation}\label{gapheight}
f({(T/2)}^+;T) - f({(T/2)}^-;T) = 2\psi(T/2)\varphi(T/2).
\end{equation} 
Here $(T/2)^\pm$ means we approach $T/2$ from left/right. When $T\to\infty$ the gap closes, i.e. tends to zero. This prediction is later verified for the simulations presented in Fig. \ref{fig:twolimits}. Similarly for $N\ge 3$ the maximum has to be larger than $T/N$. So the PDF $f(m;T)$ is non-analytical at points $T/N$ which all are in the first half time $m<T/2$, see \cite{godreche2015statistics}. 

\subsection{Long-time limits for fractal renewal processes}

We calculate the long-time limit $T\to\infty$ of $f(m;T)$ in the second half $T/2< m< T$, i.e. Eq.~(\ref{formal0}), for power law waiting time PDFs
\begin{equation}\label{plw}
\psi(\tau)\sim b_\alpha\tau^{-1-\alpha}
\end{equation}
with $\alpha\in(0,1)$ or $\alpha\in(1,2)$. The second half implies that we are dealing with large values of $m$. Therefore we consider the linear order $m=\mathcal{O}(T)$ when the maximum is of the order of the observation time. Then the second half shows rich behavior for $f(m;T)$ for the power law waiting time PDF Eq.~(\ref{plw}) as described now.\\

\begin{figure}[H]
  \centering\includegraphics[width=1\linewidth]{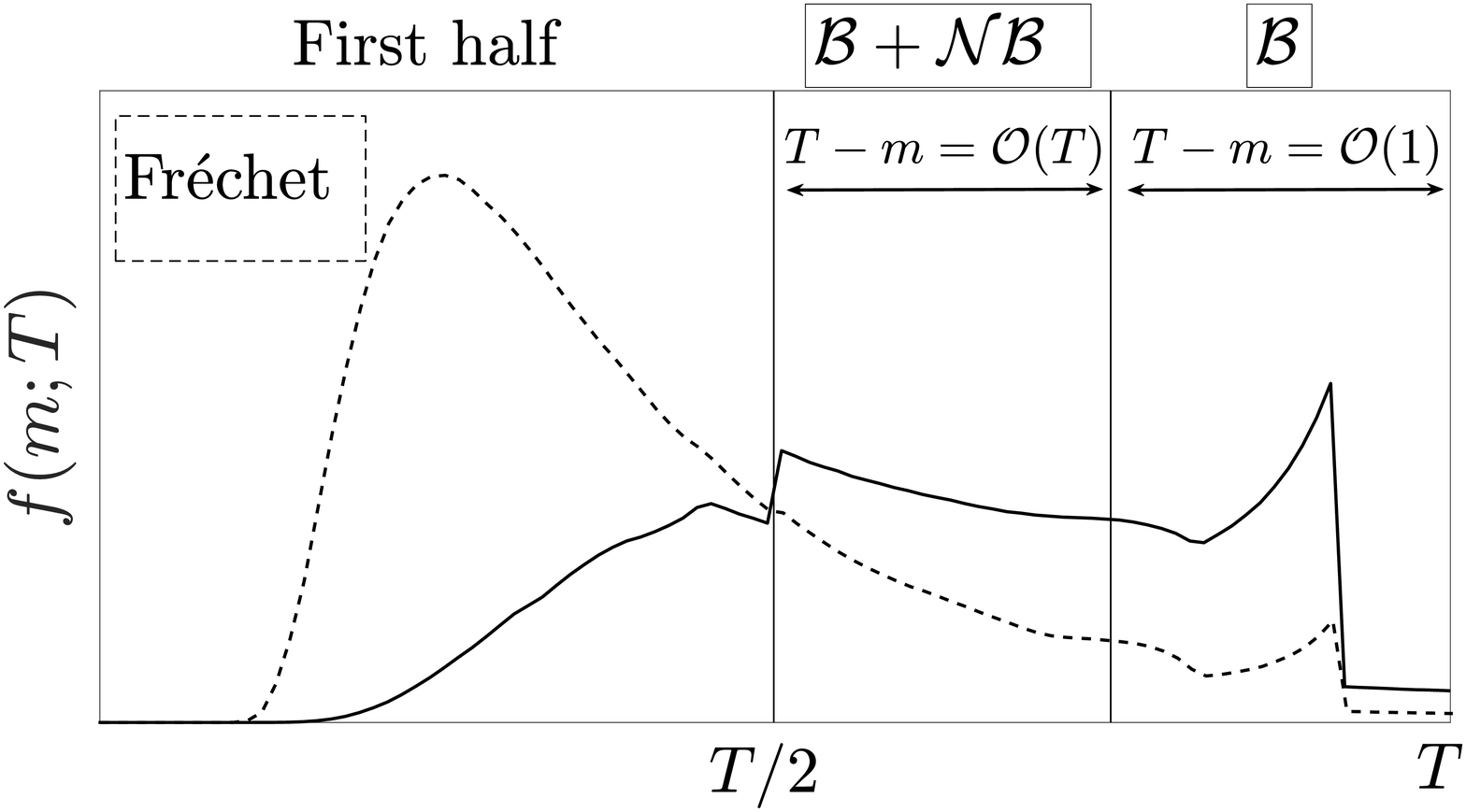}
\caption{\small{Presentation of the two scaling regimes in the second half $T/2<m< T$ which we apply for power law waiting time PDFs with $\alpha\in(0,1)$ (black line) and $\alpha\in(1,2)$ (dashed line). When the remaining time scales linearly $T-m=\mathcal{O}(T)$ both processes $\mathcal{B}$ and $\mathcal{NB}$ contribute. We obtain the scaling function $\mathcal{G}(m/T)$ in Eq.~(\ref{g}) for $\alpha\in(0,1)$ and $\mathcal{I}(m/T)$ in Eq.~(\ref{infinite}) for $\alpha\in(1,2)$. When $T-m=\mathcal{O}(1)$ then only $\mathcal{B}$ contributes which is related to the rate function, see Eq.~(\ref{secondkind22}). For $\alpha\in(0,1)$, there is a matching between $\mathcal{G}$ and the $T-m=\mathcal{O}(1)$ regime but not between the first half time and $\mathcal{G}$ because $f(m;T/2)$ is not differentiable. For $\alpha\in(1,2)$, there is a matching between Fr\'echet's law and $\mathcal{I}$ and also between $\mathcal{I}$ and the $T-m=\mathcal{O}(1)$ regime. For a detailed analysis of small $m$ we refer to \cite{godreche2015statistics}.}}\label{fig:tworegimes}
\end{figure}

 Since $R(T-m)$ and $\langle N(T-m)\rangle$ depend on the remaining time $T-m$ we have to specify how the remaining time $T-m$ behaves. We first consider linear order $T-m=\mathcal{O}(T)$, see Fig.~\ref{fig:tworegimes}. Hence we have to calculate the long-time limit of $R$ and $\langle N \rangle$. This is equivalent to calculating the small $s$ behavior of Eq.~(\ref{largetau}). The small $s$ behavior of the waiting time PDF is
\begin{equation}\label{smalls}
\widehat{\psi}(s) \sim 
\begin{cases}
1- b_\alpha |\Gamma(-\alpha)|s^\alpha & \text{ for } \alpha\in(0,1),\\
1- \langle\tau  \rangle s & \text{ for } \alpha\in(1,2)
\end{cases}
\end{equation}
with the mean waiting time $\langle \tau \rangle=\int_0^\infty \tau\psi(\tau)d\tau$.\\

\begin{figure}[H]
  \centering\includegraphics[width=1\linewidth]{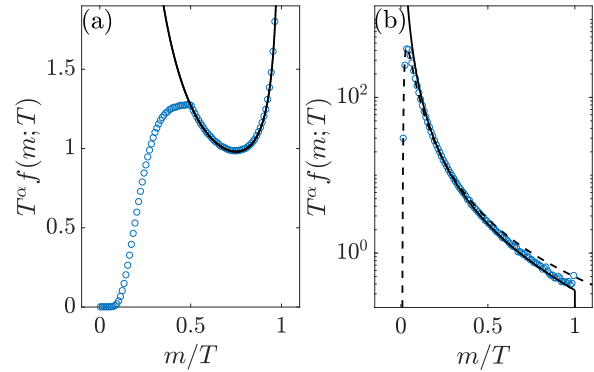}
\caption{Rescaled histogram from simulations (blue circles) for the Pareto waiting time PDF with (a) $\alpha=1/2$ and (b) $\alpha=3/2$ compared with
the theory (black line), i.e. (a) $\mathcal{G}$ of Eq.~(\ref{g}) and (b) $\mathcal{I}$ of Eq.~(\ref{infinite}). In the latter, we also plotted Fr\'echet's law Eq.~(\ref{typ2}) (dashed line). The number of realizations is $5\times 10^7$. The observation time is (a) $T=5000$ and (b) $T=1000$. Note that when $T\to\infty$ the discontinuity for $\alpha\in(0,1)$ is with respect to the derivation of $f_1(m;T)$ at $T/2$, while for finite time $T$ we observe a gap discontinuity of the maximum PDF itself, see Fig.~\ref{fig:four}.}\label{fig:twolimits}
\end{figure}

 In case of $\alpha\in(0,1)$ we obtain from the small $s$ behavior of Eq.~(\ref{largetau}) the scaling law of the second half maximum PDF as
 \begin{equation}\label{firstg}
 f(m;T) \sim \frac{1}{T} \mathcal{G}\left(\frac{m}{T}\right)
 \end{equation}
 with
\begin{equation}\label{g}
\mathcal{G}\left(\xi\right) = \frac{\text{sin}(\pi \alpha)}{\pi}{\xi}^{-\alpha}(1-\xi)^{\alpha-1} + \frac{\text{sin}(\pi \alpha)}{\pi} \xi^{-\alpha-1}(1-\xi)^\alpha.
\end{equation} 
The rescaled variable is $\xi=m/T$. The first term represents processes $\mathcal{B}$ and the second term $\mathcal{NB}$. Of course one can sum both terms and get the right hand side as $\text{sin}(\pi \alpha)/\pi \xi^{-1-\alpha}(1-\xi)^{\alpha-1}$ which was already found in \cite{godreche2015statistics,
lamperti1961contribution}. This function is valid for $1/2 < \xi < 1$ due to the restriction on the second half. The mid-point $\xi=1/2$ is non-analytical as it can be seen by the kink in Fig.~\ref{fig:twolimits} where we compare the theory with Monte Carlo simulations. Note that for $\xi\to 1$ the function blows up to infinity. In reality for any finite observation time, the maximum PDF does not diverge. 
Hence later we cure this problem by considering constant remaining time $T-m=\mathcal{O}(1)$, see also Fig.~\ref{fig:tworegimes}. That analysis will show how a second scaling law describes rare events. \\

\begin{center}
\begin{figure*}
  \centering\includegraphics[width=0.8\linewidth]{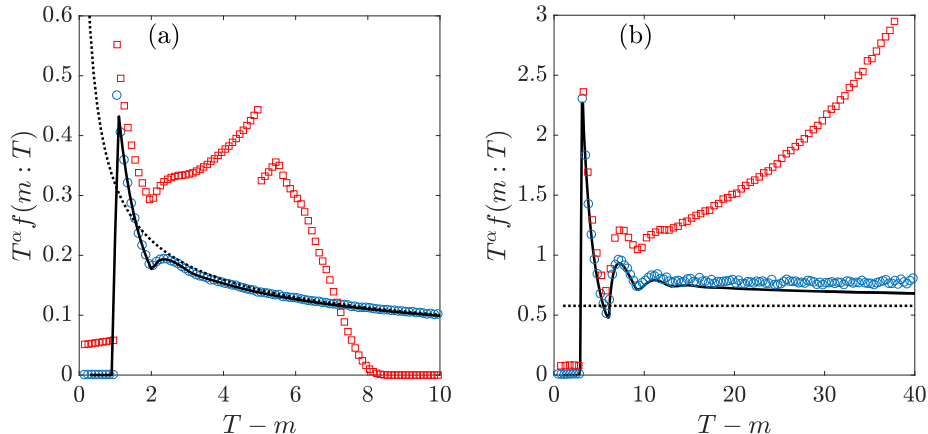}
\caption{Rescaled histogram from simulations (red squares and blue circles) compared with the theory $b_\alpha/\alpha R(T-m)$ of Eq.~(\ref{secondkind22}) (black line) and the matching functions of (a) Eq.~(\ref{harmonized1}) and (b) Eq.~(\ref{harmonized2}) (dotted lines). The simulations were performed for two different times in each figure: (a) $T=10$ (red squares) and $T=1000$ (blue circles), and (b) $T=100$ (red squares) and $T=1000$ (blue circles). The number of realizations is (a) $10^8$ and (b) $5\times 10^7$. Note that we used $\tau_0=3$ and not $\tau_0=1$ in (b) to reduce computation time.}\label{fig:aglet}
\end{figure*}
\end{center}

In case of $\alpha\in(1,2)$ we obtain from the small $s$ behavior of Eq.~(\ref{largetau}) the scaling law of the second half maximum PDF as
\begin{equation}\label{frechetcure}
f(m;T) \sim \frac{1}{T^\alpha} \mathcal{I}\left(\frac{m}{T}\right)
\end{equation}
with
\begin{equation}\label{infinite}
\mathcal{I}(\xi) = \frac{b_\alpha}{\langle\tau\rangle} \xi^{-1-\alpha}(1-\xi)+ \frac{b_\alpha}{\alpha\langle\tau\rangle}\xi^{-\alpha}.
\end{equation}
The rescaled variable is $\xi=m/T$. The first term represents processes $\mathcal{B}$ and the second term $\mathcal{NB}$. This formula was already found in the context of the big jump principle in physical modelling for the two-state L\'evy walk \cite{vezzani2019single,
wang2019transport,
vezzani2020rare,burioni2020rare}. The important point of this scaling law is that it cures the non-physical diverging second moment of Fr\'echet's law as pointed out by \cite{vezzani2019single}. Simply put, most values of the random variable $\tau_\text{max}$ are found for values below $m<T/2$ when $T$ is large. These typical events follow Fr\'echet's law
\begin{equation}\label{typ2}
f(m;T)\sim \frac{1}{(T/\langle \tau \rangle)^{1/\alpha} } b_\alpha \xi^{-1-\alpha} \text{exp}\left(-\frac{b_\alpha \xi^{-\alpha}}{\alpha}\right).
\end{equation}
with $\xi=m (T/\langle \tau \rangle)^{-1/\alpha}$, i.e. $m=\mathcal{O}(T^{1/\alpha})$. See \cite{godreche2015statistics} for a rigorous derivation. But Fr\'echet's law predicts the divergence of the variance of $\tau_\text{max}$ which is non-physical since $m\le T$. The scaling law Eq.~({\ref{typ2}) matches with Fr\'echet's law: The small $m$ behavior of Eq.~(\ref{frechetcure}) equals the large $m$ behavior of Fr\'echet's law Eq.~(\ref{typ2}), namely $b_\alpha Tm^{-1-\alpha}/\langle \tau \rangle$. Hence both scaling regimes, i.e. Fr\'echet and the far tail, are complementary. In Fig.~\ref{fig:twolimits} where we compare Eq.~(\ref{frechetcure}) with numerical simulations and Fr\'echet's law Eq.~(\ref{typ2}). We see that the gap at the mid vanishes. Furthermore as explained in \cite{vezzani2019single,
wang2019transport,
rebenshtok2014non}, the function $\mathcal{I}$ of Eq.~({\ref{infinite}) is an infinite covariant density because it is non-normalizable. However, it describes the second moment of $f(m;T)$. \\

Summarized, although the limit laws Eq.~(\ref{firstg}) and (\ref{frechetcure}) are known we showed that they arise from the second half distribution Eq.~(\ref{formal0}). It is not surprising for $\alpha\in(0,1)$ but for $\alpha\in(1,2)$ because of the different behavior of the non-analytical midpoint in the thermodynamic limit. We now present a new result unravelled by our approach.

\subsection{Long-time limits for fractal renewal processes with constant remaining time}\label{sec:aglet}

Here we calculate the long-time limit of $f(m;T)$ in the second half time $T/2 < m < T$, i.e. Eq.~(\ref{formal0}), also with $m=\mathcal{O}(T)$ but now we consider constant remaining time $T-m=\mathcal{O}(1)$, see Fig. \ref{fig:tworegimes}. Hence the rate function $R(T-m)$ and the mean number of events $\langle N(T-m)\rangle$ stay constant in Eq.~(\ref{formal0}). So we only have to compare their prefactors $\varphi(m)\sim b_\alpha m^{-\alpha}/\alpha$ and $\psi(m)\sim b_\alpha m^{-1-\alpha}$. The first one is dominant and scales as $b_\alpha T^{-\alpha}/\alpha$. The second term can be neglected. Therefore we find the scaling law
\begin{equation}\label{secondkind22}
f(m;T) \sim  \frac{1}{T^\alpha}\frac{b_\alpha}{\alpha} R(T-m)
\end{equation}
which is valid for both cases $\alpha\in(0,1)$ and $\alpha\in(1,2)$. This formula means that the maximum waiting time is always the last one, i.e. this long-time limit comes solely from the process $\mathcal{B}$. If a waiting time is the maximum but not the last one then it has to end exactly in such a way that the remaining time $T-m$ is of order $1$. But as $T$ increases this probability becomes zero so that only $\mathcal{B}$ contributes. Furthermore, the remaining time can be a small value and therefore the full form of the waiting time PDF $\psi(\tau)$ (and consequently the full form of $R(T-m)$) is required. This is in contrast to the previous study of $T-m=\mathcal{O}(T)$ where the asymptotic behavior of $\psi(\tau)$ fully describes the scaling of the maximum PDF in the second half.\\ \par

The scaling function $b_\alpha R(T-M)/\alpha$ is obviously non-normalisable because
\begin{equation}
T^\alpha \int_0^T f(m;T)dm\sim\frac{b_\alpha}{\alpha}\int_0^\infty R(\epsilon) d\epsilon  \to \infty
\end{equation}
with $\epsilon=T-m$. Technically, the integral over $m$ shown here is only correct for $m>T/2$ but this doesn't change the divergence. However, $b_\alpha R(T-m)/\alpha$ matches with integrable scaling PDFs: For $\alpha\in(0,1)$ it matches with $\mathcal{G}$ and for $\alpha\in(1,2)$ it matches with $\mathcal{I}$ which matches itself with Fr\'echet's law. The small $T-m$ limit of $f(m;T)$ with $T-m=\mathcal{O}(T)$ is the same as the large $T-m$ limit of $f(m;T)$ with $T-m=\mathcal{O}(1)$. For $\alpha\in(0,1)$ this is
\begin{equation}\label{harmonized1}
f(m;T) \sim \frac{\text{sin}(\pi\alpha)}{\pi} T^{-\alpha} (T-m)^{\alpha-1}
\end{equation}
and for $\alpha\in(1,2)$ this is
\begin{equation}\label{harmonized2}
f(m;T) \sim \frac{b_\alpha}{\alpha \langle\tau \rangle} T^{-\alpha} .
\end{equation}
In Fig. \ref{fig:aglet} we simulate $f(m;T)$ for two different values of $\alpha$ and compare $T^\alpha f(m;T)$ plotted over $T-m$ with $b_\alpha/\alpha R(T-m)$ of Eq.~(\ref{secondkind22}) and also with the matching functions of Eq.~(\ref{harmonized1}) and Eq.~(\ref{harmonized2}). We find that simulation and theory matches. The difference between Eq.~(\ref{secondkind22}) and the matching functions of Eq.~(\ref{harmonized1}) and Eq.~(\ref{harmonized2}) is easy to see only when $T-m$ is relatively small. So the analysis of the $T-m=\mathcal{O}(1)$ regime is suitable to describe rare events very close to $T$.

\section{Zero range process}\label{sec:zrp}

\subsection{Basics}

Zero range processes in equilibrium describe a system with a fixed number $K$ of interacting particles. These particles are located in well separated traps where transition times between the traps are very fast. We have $N$ such traps, and in each trap $i\in[1,N]$ we have $\kappa_i\ge 0$ particles. Clearly the constraint is 
\begin{equation}
K=\sum_{i=1}^N \kappa_i,
\end{equation}
see Fig.~\ref{fig:modelconc} and Table~\ref{tab:table01}. Here $\psi(\kappa)$ is the probability of finding $\kappa_i$ particles in the trap $i$. In thermal equilibrium, $\psi(\kappa)$ is the Boltzmann factor, though more generally it depends on the microscopical description of the transitions \cite{bar2014mixed,bar2014mixed2}. In this model the number of traps $N$ is fixed, unlike the random number of renewals in the previous model. A well-studied phenomenon in this model is condensation \cite{evans2006canonical,majumdar2010real,majumdar2005nature,zia2004construction,evans2005nonequilibrium}. When the density of the system $K/N$ crosses a critical value, a macroscopic number of particles may occupy one trap. It is then natural to wonder what is the distribution $f(m;K)$ of the maximum $\kappa_\text{max}=\text{max}(\kappa_1,\ldots,\kappa_N)$ with value $m$ since that describes the statistical properties of the condensation \cite{MAJUMDAR20201,
majumdar2010real,
evans2008condensation}.

\subsection{Extreme value statistics in the second half}

We investigate the statistics of the maximum particle number \cite{MAJUMDAR20201,
majumdar2010real,
evans2008condensation}
\begin{equation}
\kappa_\text{max} = \text{max}(\kappa_1,\kappa_2,\ldots,\kappa_N).
\end{equation}
The maximum probability mass function (PMF) is defined by $f_N(m;K)=F_N(m;K)-F_N(m-1;K)$. The maximum CDF were derived in \cite{evans2008condensation} and is given by
\begin{equation}\begin{split}\label{cdfzrp}
F_N(m;K) = \frac{1}{Z_N(K)}\sum_{\bm{\kappa}=0}^m \prod_{i=1}^N \psi(\kappa_i) \delta_{K,\|\bm{\kappa}\|_1}.
\end{split}\end{equation}
Similar to Eq.~(\ref{singlecum2}) (but there for integrals) we just wrote the $N$-multiple sums shortly as $\sum_{\bm{\kappa}=0}^m=\sum_{\kappa_1=0}^m\ldots\sum_{\kappa_1=0}^m $ with the $N$-vector $\bm{\tau}=(\tau_1,\ldots,\tau_N)^\text{T}$. The partition function is the $N$-fold convolution
\begin{equation}
Z_N(K) = (\psi\ast\ldots\ast\psi)^{(N)}(K),
\end{equation}
see \cite{evans2008condensation}. The $2$-fold convolution for discrete functions is $(\psi\ast \psi)^{(2)}(K)=\sum_{\kappa_1=0}^K \psi(\kappa_1) \psi(K-\kappa_1)$ and higher orders are defined successively. Eq.~(\ref{cdfzrp}) is easy to interpret, the set of particle numbers $\{\kappa_1,\ldots,\kappa_N\}$ are all less than or equal to $m$, and the Kronecker delta is the constraint. \\ 

Similar to the \textit{RP} we can calculate the maximum PMF for the second half $K/2< m < K$ with the almost identical analysis. The maximum PMF is generally derived from Eq.~(\ref{cdfzrp}) as
\begin{equation}\label{maxpmf}
f_N(m;K) = \frac{N\psi(m)}{Z_N(K)} \sum_{\bm{\kappa}=0}^m \prod_{i=1}^{N-1} \psi(\kappa_i) \delta_{K-m,\|\bm{\kappa}\|_1}.
\end{equation}
Here $\bm{\kappa}=(\kappa_1,\ldots,\kappa_{N-1})^\text{T}$. This formula is similar Eq.~(\ref{fint}), i.e. the maximum PDF for \textit{RP} in the second half, but the integrals are replaced by sums and the delta function is replaced by the Kronecker delta. Also the \textit{ZRP} formula has only term because here exist no $\mathcal{B}$. In Appendix \ref{sec:appa} we show that the sums in Eq.~(\ref{maxpmf}) are identical to the $N$-fold convolution in the range $K/2<m<K$. This means that the maximum PMF in the second half is
\begin{equation}\label{zrppmf}
\boxed{f_N(m;K) = \frac{1}{Z_N(K)}N \psi(m)\Phi_{N-1}(K-m).}
\end{equation}
Here, we introduced 
\begin{equation}
\Phi_{N-1}(K-m) = (\psi\ast\ldots\ast\psi)^{(N-1)}(K-m)
\end{equation}
which is the PMF with value $K-m$ of the sum of $N-1$ discrete IID random variables whose common PMF is $\psi(\kappa)$. So Eq.~(\ref{zrppmf}) relates extreme statistics with one of the most well-studied problems in stochastic theory: the sum of IID random variables, in physics this is simply the problem of a $N-1$ step random walk. In addition, we see here a useful modification of the classical EVT case Eq.~(\ref{evt}). The CDF $\Psi^{N-1}(m)$ is replaced by $\Phi_{N-1}(K-m)$ which is also divided by $Z_N(K)$. When the maximum particle number $\kappa_\text{max}$ is $m$, all other particle numbers add up to the remaining number $K-m$ due to the constraint. In Fig.~\ref{fig:zrpfig} we compare theory and simulation.

\begin{figure}[H]
  \centering\includegraphics[width=0.9\linewidth]{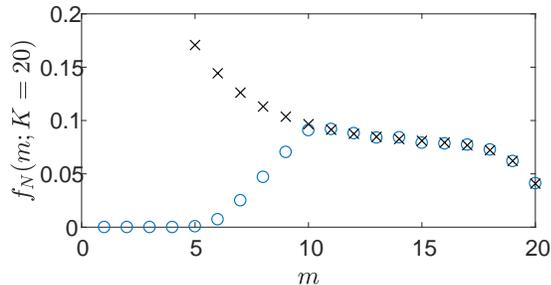}
\caption{Histogram of the maximum PMF $f_N(m;K)$ of \textit{ZRP} from Monte Carlo simulations (blue circles) compared with the theory of Eq.~(\ref{zrppmf}) (black crosses) for Zeta $\psi(\kappa)=1/\zeta(1+\alpha) (\kappa+1)^{-1-\alpha}$ with $\alpha=1/2$, $K=20$ and $N=5$. The simulations were performed with $10^7$ realizations. The analytical expression of the PMF of the sum of IID random variables $\Phi_{N-1}(K-m)$ (as well as $Z_N(K)$) is obtained via inverse $z$-transform. The kink at the midpoint $K/2$ is visible although the random variables are discrete.}\label{fig:zrpfig}
\end{figure}

\subsection{Relationship to condensation}
From the joint PMF of the particle numbers
\begin{equation}
p_N(\bm{\kappa};K)=\frac{1}{Z_N(K)} \psi(\kappa_1)\ldots \psi(\kappa_N)\delta_{K,\|\bm{\kappa}\|_1}
\end{equation}
we obtain the well-studied single trap distribution of the particle number
\begin{equation}
\rho_N(m;K)=  \frac{1}{Z_N(K)} \psi(m)\Phi_{N-1}(K-m)
\end{equation}
by summing over $N-1$ random variables, see \cite{evans2008condensation}. Comparing with Eq.~(\ref{zrppmf}) yields 
\begin{equation}\label{zrp2}
f_N(m;K) = N \rho_N(m;K)
\end{equation}
for the second half $K/2<m<K$. This is a modification of Eq.~(\ref{evt}) when we set $\Psi(m)=1$ due to the constraint. This result was obtained in \cite{majumdar2010real} as a limiting law in the condensation phase of the model. Our result shows that it is exactly valid close to and far from the thermodynamic limit, regardless of the occurrence of condensation. It is independent of the structure of $\psi(\kappa)$. Hence, our result provides a general connection between EVT and the single trap distribution of the particle number. We refer to \cite{MAJUMDAR20201,
majumdar2010real,
evans2008condensation} where the thermodynamic limit of $\rho_N(m;K)$ was studied.\\ 

Finally, the main interest of our result on the \textit{ZRP} is to demonstrate how the decoupling trick of Eq.~(\ref{convotrick}) used for the \textit{RP} can easily be applied also here. The model differences summarized in Table \ref{tab:table01} don't alter the general theme. The same is true for the last studied model.

\section{Truncated inverse distance squared Ising model}\label{sec:tidsi}
\subsection{Basics}

The \textit{TIDSI} describes a one-dimensional system of spin domains with each domain having spins $+1$ or $-1$, see Fig.~\ref{fig:modelconc}. There is an inverse squared long-range interaction between spins within the same domain. Let $N$ be the random number of domains $i\in[1,N]$ with each the domain length $\lambda_i\ge 1$. The constraint is the fixed total length of the system
\begin{equation}
L=\sum_{i=1}^N \lambda_i,
\end{equation}
see Fig.~\ref{fig:modelconc} and Table~\ref{tab:table01}. The domain $i$ of length $\lambda_i$ is associated with the weight $\psi(\lambda) \propto \lambda^{-\gamma}$ where the domain length decays with the parameter $\gamma\ge 1$ which is the product of the inverse temperature $1/(k_BT)$ and the long-range interaction \cite{bar2016exact}. The relevance of \textit{TIDSI} is that it exhibits a mixed order phase transition, i.e. it shows features of phase transitions of first and of second kind. Depending on the temperature, there is either a ferromagnetic phase with a large number of domains or a paramagnetic phase with one domain of order $L$. Thus the analysis of the extreme domain size $\lambda_\text{max}=\text{max}(\lambda_1,\ldots,\lambda_N)$ is important \cite{MAJUMDAR20201,bar2016exact}.

\subsection{Extreme value statistics in the second half}

We investigate the statistics of the maximum domain length \cite{bar2016exact}
\begin{equation}
\lambda_\text{max} = \text{max}(\lambda_1,\lambda_2,\ldots,\lambda_N).
\end{equation}
Since the number of events $N$ is random it is instructive to consider
\begin{equation}\label{sumcum}
f(m;L) = \sum_{N=1}^\infty f_N(m;L)
\end{equation}
with $f_N(m;L)=F_N(m;L)-F_N(m-1;L)$ being the maximum PDF with exactly $N$ renewal events. In this context the value of $N$ is a sampled value. The maximum CDF with given $N$ were derived in \cite{bar2016exact} and is given by
\begin{equation}
F_N(m;L) = \frac{1}{Z(L)}\sum_{\bm{\lambda}=0}^m \prod_{i=1}^N \psi(\lambda_i) \delta_{L,\|\bm{\lambda}\|_1}.
\end{equation}
This formula is almost identical to Eq.~(\ref{cdfzrp}) for \textit{ZRP} but the partition function is here
\begin{equation}
Z(L)= \sum_{N=1}^\infty (\psi\ast\ldots\ast\psi)^{(N)}(L).
\end{equation}
The maximum PMF with given $N$ is 
\begin{equation}
f_N(m;L) = \frac{N\psi(m)}{Z(L)} \sum_{\bm{\lambda}=0}^m \prod_{i=1}^{N-1} \psi(\lambda_i) \delta_{L-m,\|\bm{\lambda}\|_1}.
\end{equation}
Again, we use that in the second half $L/2<m<L$ this formula is identical to the convolution. The second half maximum PMF with given $N$ is
\begin{equation}
f_N(m;L) = N \psi(m) P_{N-1}(L-m).
\end{equation}
Here, the probability of having $N-1$ spin domains is
\begin{equation}\begin{split}
P_{N-1}(L-m) &=  \frac{1}{Z(L)} \left\langle \delta_{L-m,\|\bm{\lambda}\|_1} \right\rangle\\
&= \frac{1}{Z(L)} (\psi\ast\ldots\ast\psi)^{(N-1)}(L-m)
\end{split}\end{equation}
where the average $\langle \circ \rangle$ is performed over all possible domain lengths. Finally, averaging over all $N$ yields the second half maximum PMF
\begin{equation}\label{tidsiresult}
\boxed{f(m;L) = \psi(m) \frac{Z(L-m)}{Z(L)} (\langle N(L-m)\rangle +1 )}
\end{equation}
with $L/2<m<L$. The mean number of domains is
\begin{equation}
\langle N(L-m)\rangle = \sum_{N=1}^\infty N P_N(L-m)
\end{equation}
One could write Eq.~(\ref{tidsiresult}) also as
\begin{equation}\label{tidsiresult2}
f_N(m;L) = \frac{1}{Z(L)} \psi(m)\sum_{N=1}^\infty N \Phi_{N-1}(L-m).
\end{equation}
in order to emphasize the relationship to the random walk picture similar as we did for the \textit{ZRP}. When the maximum spin domain length $\lambda_\text{max}$ is $m$, all other lengths add up to the remaining length $L-m$ due to the constraint. In Fig.~\ref{fig:tidsifig} we compare theory and simulation.

\begin{figure}[H]
  \centering\includegraphics[width=0.9\linewidth]{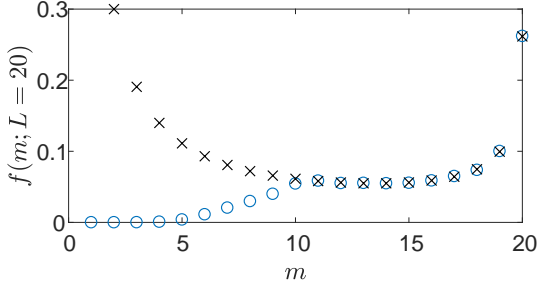}
\caption{Histogram of the maximum PMF $f(m;L)$ of \textit{TIDSI} from Monte Carlo simulations (blue circles) compared with the theory of Eq.~(\ref{tidsiresult}) (black line) for Zeta $\psi(\lambda)=1/\zeta(1+\alpha) \lambda^{-1-\alpha}$ with $\alpha=0.2$ and $K=20$. The simulations were performed with $10^6$ realizations. The analytical expression of the mean number of domains $\langle N(L-m)\rangle$ (as well as $Z(L-m)$ and $Z(L)$) is obtained via inverse $z$-transform. The kink at the midpoint $L/2$ is visible although the random variables are discrete.}\label{fig:tidsifig}
\end{figure}

\subsection{Limiting law in the critical phase}
We consider the large total length limit $L\to\infty$ of $f(m;L)$ in the second half $L/2<m<L$, i.e. Eq.~(\ref{tidsiresult}) or (\ref{tidsiresult2}), for the critical phase between ferromagnetic and paramagnetic phases \cite{bar2016exact}. Then the calculations can be transferred almost effortlessly from above \textit{RP} techniques. For the limiting laws in the ferromagnetic and paramagnetic phases we refer to \cite{bar2016exact}. The weight is generally
\begin{equation}
\psi(\lambda) = \frac{\text{e}^{-\beta\Delta}}{\lambda^{1+\alpha}}
\end{equation}
with the inverse temperature $\beta=1/(k_B T)$, the chemical potential $\Delta$ and $1+\alpha=\beta J\ge 1$ where $J$ is the strength of the inverse squared long-range interaction within a single spin domain, see \cite{bar2016exact}. In the critical phase the marginal domain size decays algebraically. Then the weight is
\begin{equation}\label{tidwait}
\psi(\lambda) = \frac{1}{\zeta(1+\alpha)\lambda^{1+\alpha}}
\end{equation}
with the Riemann Zeta function $\zeta(1+\alpha)=\sum_{N=1}^\infty N^{-1-\alpha}$, i.e. the fugacity is $\text{e}^{-\beta\Delta}=1/\zeta(1+\alpha)$. It was shown in \cite{bar2016exact} that there are two regimes in the critical phase for $\alpha\in(0,1)$ and $\alpha>1$. We restrict the latter to $\alpha\in(1,2)$ in order to compare it to \textit{RP}. As explained in \cite{bar2016exact} the analysis using $z$-transform can be replaced by Laplace transforms in the critical phase which we use now.\\

The $z$-transform of the weight is
\begin{equation}\label{ztrafo}
\psi(\lambda) \laplace \hat{\psi}(z) = \sum_{L=1}^\infty \psi(L)z^L,
\end{equation}
the $z$-transform of the denominator of Eq.~(\ref{tidsiresult}) is
\begin{equation}\begin{split}
\sum_{N=1}^\infty N \Phi_{N-1}(L-m) &\laplace \sum_{N=1}^\infty N z^m \hat{\psi}^{N-1}(z) \\
&= \frac{z^m}{\left[1-\hat{\psi}(z)\right]^2}
\end{split}\end{equation}
and the $z$-transform of the numerator of Eq.~(\ref{tidsiresult}) is
\begin{equation}\label{largez}
\sum_{N=1}^\infty  \Phi_N(L) \laplace = \sum_{N=1}^\infty \hat{\psi}^n(z) = \frac{\hat{\psi}(z)}{1-\hat{\psi}(z)}.
\end{equation}
The symbol $\laplace$ means we perform the $z$-transform as defined in Eq.~(\ref{ztrafo}).\\

We study here the scaling $m=\mathcal{O}(L)$ and $L-m=\mathcal{O}(L)$. Hence, we need the large $L$ limit of both the denominator and numerator. We set $z=\text{exp}(-s)$ and consider the small $s$-behavior of the weights
\begin{equation}
\hat{\psi}(s) \sim 
\begin{cases}
1-\frac{|\Gamma(-\alpha)|}{\zeta(1+\alpha)}s^\alpha & \text{ for } 0<\alpha<1,\\
1-\langle \lambda \rangle s & \text{ for } 1<\alpha<2.
\end{cases}
\end{equation}
This is equivalent to the asymptotic behavior of $\hat{\psi}(z)\sim 1- |\Gamma(-\alpha)|/\zeta(1+\alpha) (1-z)^\alpha -\zeta(\alpha)/\zeta(\alpha) (1-z)$ at the branch point $z=1$, see \cite{bar2016exact}.\\

For $\alpha\in(0,1)$ we get from the inverse Laplace transform the scaling law
\begin{equation}\label{our}
f(m;L) \sim \frac{1}{L} \mathcal{G}\left(\frac{m}{L}\right)
\end{equation}
with
\begin{equation}
 \mathcal{G}(\xi)=\frac{\Gamma(\alpha)}{|\Gamma(-\alpha)|\Gamma(2\alpha)}\xi^{-1-\alpha}(1-\xi)^{2\alpha-1}
\end{equation}
with the rescaled variable $\xi=m/L$. The same limiting law has been derived in \cite{godreche2017longestaa}. It has also been derived in \cite{bar2016exact} but with a different expression depending on hypergeometric functions. The results are identical, see Appendix \ref{sec:appb}. Eq.~(\ref{our}) is valid for $1/2 < \xi < 1$ due to the restriction on the second half. The midpoint $\xi=1/2$ is non-analytical as reported in \cite{bar2016exact,godreche2017longestaa}. Note that at $\xi\to 1$ the function blows up to infinity. In reality for any finite observation time, the maximum PMF does not diverge. Below we cure this problem again by considering constant remaining length $L-m=\mathcal{O}(1)$.  This describes the rare events where the scaling law Eq.~(\ref{our}) is not valid anymore.\\

 For $\alpha\in(1,2)$ we get
\begin{equation}\label{correctiontidsi}
f(m;L) \sim \frac{1}{L^\alpha} \mathcal{I}\left(\frac{m}{L}\right) 
\end{equation}
with
\begin{equation}\label{tidisiinf}
\mathcal{I}(\xi)= \frac{1}{\zeta(1+\alpha)\langle\lambda\rangle} \xi^{-1-\alpha}(1-\xi).
\end{equation}
The rescaled variable is $\xi=m/L$. The important point of this scaling law is that it cures the unphysical diverging second moment of Fr\'echet's law describing typical events
\begin{equation}\label{typ}
 f(m;L)\sim \frac{1}{(L/\langle \lambda \rangle)^{1/\alpha}} b_\alpha \xi^{-1-\alpha} \text{exp}\left(-\frac{b_\alpha \xi^{-\alpha}}{\alpha}\right)
\end{equation}
with $b_\alpha=1/\zeta(1+\alpha)$ and the rescaled variable $\xi=m (L/\langle \lambda \rangle)^{-1/\alpha}$, i.e. $m=\mathcal{O}(L^{1/\alpha})$. The mean length is $\langle\lambda \rangle = \zeta(\alpha)/\zeta(1+\alpha)$. See \cite{bar2016exact} for a rigoruous derivation. But Fr\'echet's law predicts the divergence of the variance of $\lambda_\text{max}$ which is unphysical since $m\le T$. The scaling law Eq.~({\ref{correctiontidsi}) matches with Fr\'echet's law: The small $m$ behavior of Eq.~(\ref{correctiontidsi}) equals the large $m$ behavior of Fr\'echet's law, namely $b_\alpha Lm^{-1-\alpha}/\langle \lambda \rangle$. Hence both scaling regimes are complementary. In Fig. \ref{fig:tidsifrechet} where we compare Eq.~(\ref{correctiontidsi}) with numerical simulations and Fr\'echet's law Eq.~(\ref{typ}).\\

The function of Eq.~({\ref{infinite}) is non-normalisable
\begin{equation}
L^\alpha \int_0^L f(m;L)dm\sim\frac{1}{\zeta(1+\alpha)\langle\lambda\rangle} \int\limits_0^1 \xi^{-1-\alpha}(1-\xi) d\xi \to \infty.
\end{equation}
Similar to the \textit{RP}, this limiting function describing rare events cures the infinite variance problem of Fr\'echet's law.

\begin{figure}[H]
  \centering\includegraphics[width=1\linewidth]{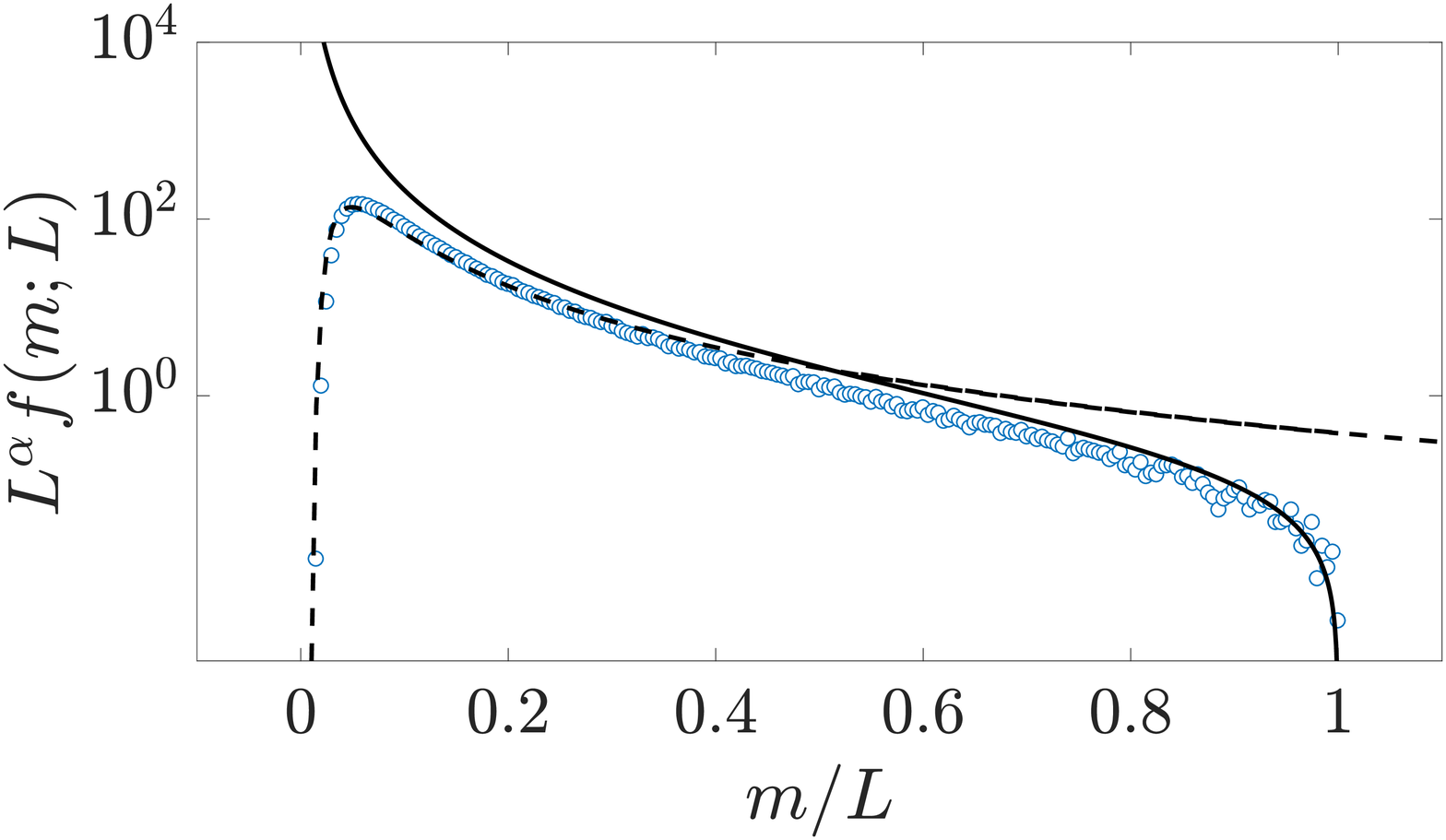}
\caption{Rescaled histogram of the maximum PMF $f(m;L)$ for \textit{TIDSI} (blue circles) for $L=200$ compared with the limiting law of Eq.~(\ref{correctiontidsi}) (solid line) and Fr\'echet's law of Eq.~(\ref{typ}) (dashed line). The simulation were performed with $10^6$ realizations and $\alpha=3/2$. Clearly, Eq.~(\ref{correctiontidsi}) works relatively well already for not too large $L$ provided that $m>L/2$.}\label{fig:tidsifrechet}
\end{figure}

\subsection{Limiting law in the critical phase with constant remaining length}

\begin{center}
\begin{figure*}
  \centering\includegraphics[width=1\linewidth]{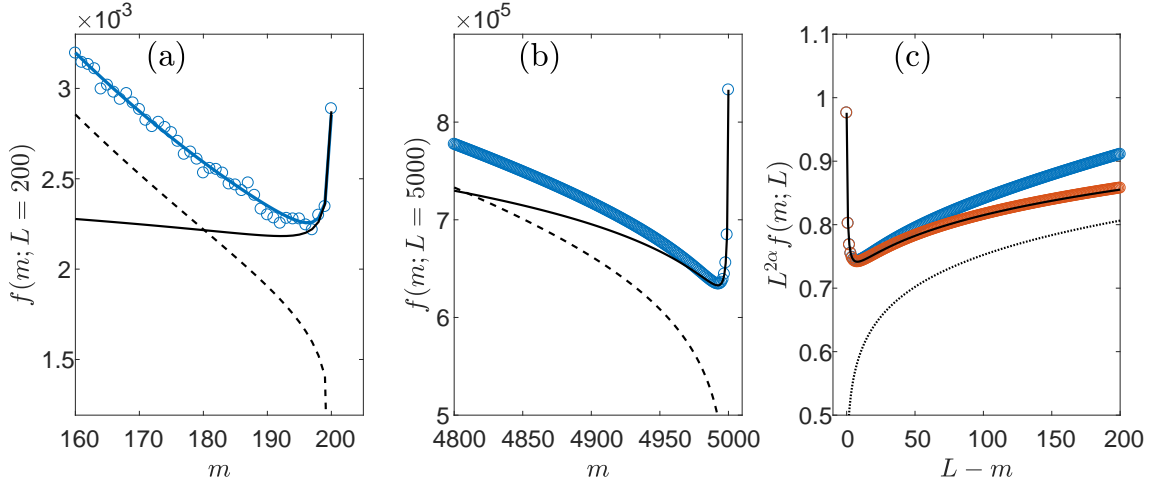}
\caption{(a) Comparison of the maximum PMF $f(m;L)$ for TIDSI with $L=200$ and $\alpha=0.55$ (see Eq.~(\ref{tidwait})) of Monte Carlo simulations (red circles), the exact half time distribution Eq.~(\ref{tidsiresult}) (red solid line), the $m/L\to \text{const.}$ scaling law Eq.~(\ref{our}) for typical fluctuations (dashed line) and the $L-m=\mathcal{O}(1)$ scaling law of Eq.~(\ref{tidsiaglet}) describing rare events (black solid line). The simulation is performed for $10^8$ realizations. (b) The maximum PMF with $L=5000$ of the exact half time distribution Eq.~(\ref{tidsiresult}) (blue circles), the $m/L\to \text{const.}$ scaling law Eq.~(\ref{our}) (dashed line) and the $L-m=\mathcal{O}(1)$ scaling law of Eq.~(\ref{tidsiaglet}) (solid line).\\
 (c) Rescaled maximum PMF plotted over $L-m$ of the exact maximum distribution Eq.~(\ref{tidsiresult}) with $L=5000$ and $L=10^5$ compared with the $L-m=\mathcal{O}(1)$ scaling law Eq.~(\ref{tidsiaglet}) and the matching function with the $L-m=\mathcal{O}(L)$ scaling law Eq.~(\ref{tidsimatch}).}\label{fig:tidsithreefigs}
\end{figure*}
\end{center}

Here, we calculate the long-time limit of $f(m;L)$ in the second half time $L/2 < m < L$, i.e. Eq.~(\ref{tidsiresult}), also with $m=\mathcal{O}(L)$ but now we consider constant remaining total length $L-m=\mathcal{O}(1)$. Hence the $Z(L-m)$ and the mean number of domains $\langle N(L-m)\rangle$ stay constant in Eq.~(\ref{tidsiresult}). So we only have to consider $\psi(m) \sim \psi(L)$ and the large $L$ behaviour of $Z(L)$. Therefore we find the scaling law
\begin{equation}\begin{split}\label{tidsiaglet}
f(m;L) &\sim \frac{\psi(L)}{Z(L)} Z(L-m) \left[ \langle N(L-m) \rangle +1\right] \\
&= Z(L-m) \left[ \langle N(L-m) \rangle +1\right]\\
&\times 
\begin{cases}
 \frac{|\Gamma(-\alpha)|\Gamma(\alpha)}{\zeta^2(1+\alpha)} L^{-2\alpha} & \text{ for } \alpha\in(0,1),\\
\frac{\langle\lambda\rangle}{\zeta(1+\alpha)}L^{-1-\alpha} & \text{ for } \alpha\in(1,2).
\end{cases}
\end{split}\end{equation}
In particular, for $Z(L)$ we used the small $z$ behaviour of Eq.~(\ref{largez}) and calculated the inverse Laplace transform with $z=\text{exp}(-s)$.\\

 The meaning of this scaling law is similar to the \textit{RP} limiting law in section \ref{sec:aglet}: it describes the rare events of $m$ very close to the constraint $L$. In Fig.~\ref{fig:tidsithreefigs}a) we show Monte Carlo simulations for a system size $L=200$. The figure illustrates that the exact expression for $f(m;L=200)$ in the second half Eq.~(\ref{tidsiresult}) work well as expected. The region near $L$ is well described by the asymptotic theory Eq.~(\ref{tidsiaglet}) while the law Eq.~(\ref{our}) is not performing well. The latter observation is to be expected as we are dealing with the rare events. Then in Fig.~\ref{fig:tidsithreefigs}b) we consider a larger system, $L=5000$. Here, Monte Carlo simulations do not converge in a reasonable time. We can however explore this regime with our exact solution, Eq.~(\ref{tidsiresult}), again a solution valid  in
the domain $m>L/2$. This points out to the fact that the exact solution can be exploited to investigate rare fluctuations where sampling of rare events, at least with straight forward simulations, is difficult or impossible. Further the exact theory also matches the asymptotic theory where it should, namely on the far right hand side of the figure. We are able to plot the exact behaviour of $f(m;L)$ near $L$ for any large value of $L$. The detailed procedure is explained as follows. First, we replace $Z(L$) by its large $L$ behaviour $\zeta(1+\alpha)/[|\Gamma(-\alpha)|\Gamma(\alpha)]L^{-1+\alpha}$ in Eq.~(\ref{tidsiresult}). Secondly, the denominator $Z(L-m)[ \langle N(L-m)\rangle+1]$ is exactly obtained via Taylor series of its $z$ transform. And here it is important that the expression $Z(L-m)[ \langle N(L-m)\rangle+1]$ only depends on $L-m$. Since we are only interested in small $L-m \leq 200$ we are able to derive the Taylor series for any value of $L$ with Mathematica. Thus, we obtain the exact expression of $f(m;L)$ near $L$. Finally, we compare this replacement of the data with the scaling laws in Fig.~\ref{fig:tidsithreefigs}. \\

The matching between the two scaling laws with $L-m=\mathcal{O}(L)$ of Eq.~(\ref{our}) and $L-m=\mathcal{O}(1)$ of Eq.~(\ref{tidsiaglet}) can be analytically calculated with an argumentation identically to the previous \textit{RP} comparison between the two regimes with $T-m=\mathcal{O}(T)$ and $T-m=\mathcal{O}(1)$ in section \ref{sec:aglet}. The small $L-m$ limit of $f(m;L)$ with $L-m=\mathcal{O}(L)$ is equal to the large $L-m$ limit of $f(m;L)$ with $T-m=\mathcal{O}(1)$. For $\alpha\in(0,1)$ this is
\begin{equation}\label{tidsimatch}
f(m;L) \sim \frac{\Gamma(\alpha)}{|\Gamma(-\alpha)|\Gamma(2\alpha)} L^{-2\alpha} \left(L-m \right)^{2\alpha-1},
\end{equation}
see Fig.~\ref{fig:tidsithreefigs}. For $\alpha\in(1,2)$ the matching function is
\begin{equation}
f(m;L) \sim \frac{1}{\zeta(1+\alpha)\langle \lambda \rangle}L^{-1-\alpha} (L-m).
\end{equation}
An interesting observation is that this matching function for different values of $\alpha\in(0,1)$ behaves totally different than the exact solution when $m\to L$, In Fig.~\ref{fig:tidsithree}, we compare for $\alpha=0.45, 0.5$ and $0.55$ the matching function Eq.~(\ref{tidsimatch}) with Eq.~(\ref{tidsiaglet}). Although both solutions match for small $m$, the $L-m=\mathcal{O}(1)$ law Eq.~(\ref{tidsiaglet}) diverges at $m\to L$ while the matching solution Eq.~(\ref{tidsimatch}) (and therefore also the $L-m=\mathcal{O}(L)$ scaling law) change its behaviour at $\alpha=1/2$. This behaviour shows that the rare event behaviour is correctly described by assuming $L-m=\mathcal{O}(1)$.

\begin{figure}[H]
  \centering\includegraphics[width=1.1\linewidth]{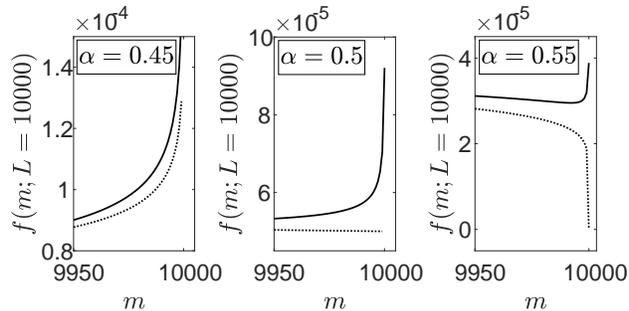}
\caption{Right tails of $f(m;L)$ for \textit{TIDSI} with $L=10^4$ described by the
$L-m=\mathcal{O}(1)$ limiting law of Eq.~(\ref{tidsiaglet}) (solid line) and the matching function to the $L-m=\mathcal{O}(L)$ limiting law Eq.~(\ref{tidsimatch}) (dotted line). Three different values of $\alpha\in(0,1)$ have been chosen. There is a significant change of the behaviour of the matching function when $\alpha$ passes $1/2$ while the exact behaviour always increases for $m\to L$.}\label{fig:tidsithree}
\end{figure}

\section{Summary}\label{sec:conclusion}

\begin{center}
\begin{table*}[]
  \begin{tabular}{ l|l|l|lrl|lrl}
  \hline
  Model & $\alpha \in$ & First half: $m\in(0,C/2)$  & \multicolumn{6}{c}{Second half: $m\in(C/2,C)$ }\\
  
  &&& \multicolumn{3}{l}{Remaining constraint $C-m=\mathcal{O}(C)$ } & \multicolumn{3}{l}{Remaining constraint $C-m=\mathcal{O}(1)$} \\ \hline
  
  \textit{RP}& $(0,1)$ & Beyond Fr\'echet's law & a) & $Tf(m;T)$ & $\sim \xi^{-1-\alpha}(1-\xi)^{\alpha-1}$ & \textcolor{red}{e)} & $T^\alpha f(m;T)$ &  $\sim  R(T-m)$ \\
  
  & $(1,2)$ & Fr\'echet's law & b) &$T^\alpha f(m;T)$ & $\sim \xi^{-1-\alpha}[1-(1/\alpha-1)\xi]$ & \textcolor{red}{f)} & $T^\alpha f(m;T)$ &  $\sim  R(T-m)$ \\ \hline

  \textit{TIDSI}& $(0,1)$ & Beyond Fr\'echet's law &c) & $Lf(m;L)$ & $\sim \xi^{-1-\alpha}(1-\xi)^{2\alpha-1}$ & \textcolor{red}{g)} &$L^{2\alpha}f(m;L)$ &  $\sim Z(L-m)[\langle N(L-m)+1] \rangle $ \\
  
& $(1,2)$ & Fr\'echet's law & \textcolor{red}{d)} &$L^\alpha f(m;L)$ & $\sim \xi^{-1-\alpha}(1-\xi)$ & \textcolor{red}{h)} &$L^{1+\alpha}f(m;L)$ &  $\sim Z(L-m)[\langle N(L-m)+1] \rangle $ \\ \hline

  \end{tabular}\caption{\label{tab:table} Collecting of limiting laws of $f(m;C)$ for \textit{RP} with $C=T$ and \textit{TIDSI} with $C=L$ in the critical phase. The random variables (waiting times and spin domains lengths) are fat-tail distributed with exponent $\alpha$. For the study on the first half $m\in(0,C/2)$ we refer for \textit{RP} to \cite{godreche2015statistics} and for \textit{TIDSI} to \cite{bar2016exact}, in particular for $\alpha\in(0,1)$ the first half shows a different scaling law than Fr\'echet's law. In the second half, i.e. $C/2<m<C$, the scaling $m=\mathcal{O}(C)$ is applied. The rescaled variable is $\xi=m/C$. We find again the laws of a) \cite{lamperti1961contribution,godreche2015statistics}, b) \cite{vezzani2019single}  and c) \cite{godreche2017longestaa}. Another expression of c) has been derived in \cite{bar2016exact} using
other methods, they express the law as a sum of two hypergeometric
funtions while the expression in the table is simpler, see Appendix \ref{sec:appb}. The limiting laws d)-h) are first presented in this article thus marked red. Note that we present the expressions in this table without prefactors.}\end{table*}
\end{center}

We have analyzed EVT of the longest waiting time $\tau_\text{max}$ of the \textit{RP}, the largest particle number per site $\kappa_\text{max}$ of the \textit{ZRP} and the largest spin domain size $\lambda_\text{max}$ of the \textit{TIDSI}. These three models share the global constraint for the sum of the random variables, i.e. the waiting times, the particle numbers per site and the spin domain lengths. The exact details of the models differ from each other. While the number of sites in the \textit{ZRP} is fixed, the number of waiting times/spin domains is random in the \text{RP}/\textit{TIDSI}. Furthermore, the last waiting time for the \textit{RP} is cut off to the backward recurrence time. However, we found that despite these differences the common trait of the global constraint enabled us to decouple the problem when the extreme value is larger than half of the constraint. One of our main results is the revelation of the deep connection between two different fields, constrained EVT and well-known quantifiers of stochastic dynamics. The latter are the mean number of renewal events Eq.~(\ref{rppdfnew}), the sum of independent and identically distributed random variables Eq.~(\ref{zrp2}) and the mean number of spin domains Eq.~(\ref{tidsiresult}). Our results are in perfect accordance in the second half of the support as presented in Fig.~\ref{fig:four} for the \textit{RP}, in Fig.~\ref{fig:zrpfig} for the \textit{ZRP} and in Fig.~\ref{fig:tidsifig} for the \textit{TIDSI}. In these figures the practical calculation of the theory relies on
Laplace transforms (or $z$ transforms). Since our theory relates two fields, namely EVT and underlying stochastic dynamics, we demonstrated exemplary for \textit{RP} in Fig.~\ref{fig:rpcdf} that one can also to obtain the EVT indirectly: The estimation of the mean number of renewals $\langle N \rangle$ is sufficient to obtain the maximum CDF by using Eq.~(\ref{rpcdfnew}). Another advantage of our theory is that we can plot the extreme value statistics for cases when Monte Carlo sampling demands huge computational resources. This was demonstrated for the \textit{TIDSI} in Fig.~\ref{fig:tidsithreefigs}.\\

After this general result of the second half maximum distribution, we considered different asymptotic limits for \textit{RP} with power law waiting times with exponent $\alpha\in(0,1) $ and $\alpha\in(1,2)$. For \textit{TIDSI} we have chosen to study the asymptotic limit in the critical phase between ferromagnetic and paramagnetic phases because the behavior is comparable to the \textit{RP} behavior. We recapped known results and also found new limiting laws when the global constraint diverges $C\to\infty$, i.e. we have diverging observation time $C=T$ for the \textit{RP} and diverging total domain length $C=L$ for the \textit{TIDSI}. The limiting behavior of the maximum distribution $f(m;C)$ is summarized as

\begin{itemize}
\item[a)] When $C\to\infty$ and $m/C$ is fixed, the second half maximum distribution for $\alpha\in(0,1)$ describes typical events. For the \textit{RP} we found Eq.~(\ref{g}) and explained previous results \cite{godreche2015statistics,lamperti1961contribution} by identifying contributions from $\mathcal{B}$ and $\mathcal{NB}$, i.e. both processes with the maximum being the last waiting time or not. For the \textit{TIDSI} we found Eq.~(\ref{our}) which was derived in \cite{godreche2017longestaa,bar2016exact}.

\item[b)]  When $C\to\infty$ and $m/C$ is fixed, the second half maximum distribution for $\alpha\in(1,2)$ complements the typical events described by Fr\'echet's law. For the \textit{RP} we found Eq.~(\ref{infinite}) which was derived in \cite{vezzani2019single}. For the \textit{TIDSI} we found Eq.~(\ref{correctiontidsi}). Both limiting laws are infinite densities.

\item[c)] We find the rare events of the statistics of the maximum for $\alpha\in(0,1)$. Especially for the \textit{RP} this is relevant because it cures the divergent behavior of the typical events near the observation time. The scaling of the rare events assumes $T-m$ is fixed while $T\to\infty$. Eq.~(\ref{secondkind22}) shows that only the process $\mathcal{B}$ is important. Here the rate function, of the mean number of renewals, is a useful tool in the analysis of the large deviations. Of course while this rate function describes rare events, it is very different from the rate function of standard large deviation theory \cite{touchette2009large}. Finally, the presented results are used in \cite{wanlifuture} where we established the so called big jump principle \cite{vezzani2019single} for the ballistic L\'evy walk model. In summary, there we show the usefulness of the approach, in the sense that the statistics of $\tau_\text{max}$ might be used to predict the large deviations of a widely applicable model of anomalous transport. In addition, we found the same scaling behaviour to describe the rare event near $L$ for the \textit{TIDSI} in the critical phase in Eq.~(\ref{tidsiaglet}). However, there is obviously no distinction between $\mathcal{B}$ and $\mathcal{NB}$ necessary.

\end{itemize}

We collect the just described limiting laws in Table \ref{tab:table} together with the behavior in the first half $0<m<T/2$. While for classical EVT the limiting behaviour is described by Fr\'echet's law, the global constraint yields rich limiting behaviour with different scaling laws for which our theory provides a helpful tool to derive them as presented in the main text.\\

\textbf{Acknowledgment} After this paper was completed C. Godr\`eche published related results \cite{godreche20202preprint}. We also thank him for pointing out Refs. \cite{wendel1964zero,godreche2017longestaa}.  M.H. is funded by the Deutsche Forschungsgemeinschaft (DFG, German Research Foundation) $–$ 436344834. E.B. acknowledges the Israel Science Foundations Grant No. 1898/17. W.W. was supported by Bar-Ilan University together with the Planning and Budgeting Committee fellowship program.

\begin{appendix}


\section{Integrals identical to the convolution}\label{sec:appa}
We first consider the \textit{RP} and later summarize the results also for the \textit{ZRP} and the \textit{TIDSI}. In Eq.~(\ref{inttest}) we have integrals of the form
\begin{equation}\label{whatisintegral}
I_N(m,T^\prime) = \int\limits_0^m d\tau_1 \ldots \int\limits_0^m d\tau_N  \prod_{i=1}^N g_i(\tau_i)\delta\left(T^\prime - \sum_{j=1}^N \tau_j\right).
\end{equation}
Note that in Eq.~(\ref{inttest}) there are $(N-1)$-multiple integrals but we consider now $N$-multiple integrals. The functions $g_i(\tau_i)$ in Eq.~(\ref{inttest}) are the waiting time PDFs $\psi(\tau_i)$ or the survival probability $\varphi(\tau_i)$. Furthermore the parameter $T^\prime$ in Eq.~(\ref{inttest}) is the remaining time $T-m$. Here we discuss general functions which must be positive $g_i(\tau_i) \ge 0$ with positive arguments $\tau_i\ge 0$. And we consider an arbitrary constraint $T^\prime>0$. The main result of this section is that the integral $I_N(m,T^\prime)$ is identical to the convolution
\begin{equation}\label{convtheorem}
I_N(m,T^\prime) =  (g_1\ast\ldots\ast g_N)^{(N)}(T^\prime)
\end{equation}
when the condition $m>T^\prime$ is fulfilled. This condition will lead to the range of the second half $T/2<m<T$ when $T^\prime=T-m$. The $2$-fold convolution is $(g_1\ast g_2)^{(2)}(T^\prime)=\int_0^{T^\prime} d\tau_1 g_1(\tau_1) g_2(T^\prime-\tau_1)$ and higher orders are defined successively.\\

We derive Eq.~(\ref{convtheorem}) with a proof by induction. Let us start with $N=2$, i.e. we show now that
\begin{equation}\label{doubleintegral}
I_2(m,T^\prime)=(g_1\ast g_2)^{(2)}(T^\prime)
\end{equation}
when $m>T^\prime$. Per definition we have
\begin{equation}\label{twotwo}
I_2(m,T^\prime)= \int\limits_0^m d\tau_1  \int\limits_0^m d\tau_2  g_1(\tau_1) g_2(\tau_2)\delta\left(T^\prime -\tau_1-\tau_2\right).
\end{equation}
For the inner integral we take both limits to infinity while putting two Heaviside functions into the integrand
\begin{equation}\begin{split}
&\int\limits_0^m d\tau_2 g_2(\tau_2) \delta(T^\prime -\tau_1-\tau_2) \\
&= \int\limits_{-\infty}^{+\infty} d\tau_2 g_2(\tau_2) \Theta(\tau_2) \Theta(m-\tau_2) \delta(T^\prime -\tau_1-\tau_2)\\
&=g_2(T^\prime -\tau_1)\Theta(T^\prime-\tau_1)\Theta(m-[T^\prime-\tau_1]).
\end{split}\end{equation}
Hence this inner integral is only nonzero under the condition
\begin{equation}\label{firstcondition}
T^\prime - m < \tau_1 < T^\prime.
\end{equation}
The further analysis of the outer integral of Eq.~(\ref{twotwo}) depends on this condition Eq.~(\ref{firstcondition}) and the relationship between $T^\prime$ and $m$. We may consider the three regimes
\begin{equation}\begin{split}\label{secondcondition}
&\text{(a) } 0<T^\prime < m, \\
&\text{(b) } m<T^\prime<2m, \\
&\text{(c) } 2m<T^\prime.
\end{split}\end{equation}
Both conditions of Eq.~(\ref{firstcondition}) and Eq.~(\ref{secondcondition}) lead to 
\begin{equation}
I_2(m,T^\prime) = 
\begin{cases}
\int\limits_0^{T^\prime} d\tau_1 g_1(\tau_1) g_2(T^\prime-\tau_1) & \text{ for (a)},\\
\int\limits_{T^\prime-m}^m d\tau_1 g_1(\tau_1) g_2(T^\prime-\tau_1) & \text{ for (b)},\\
0 & \text{ for (c)}.
\end{cases}
\end{equation}
See also Fig. \ref{fig:two} for three different areas of integration. We are only interested in the first regime when $0<T^\prime<m$. Then the double integral is the convolution and hence Eq.~(\ref{doubleintegral}) is shown for $N=2$.\\

\begin{figure}[H]
  \centering

    \includegraphics[width=0.45\textwidth]{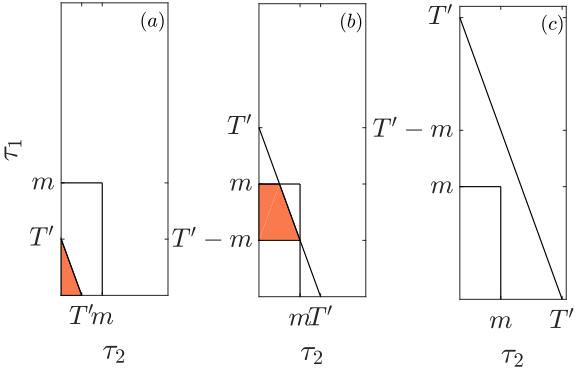}
 
  \caption{Areas of integration of $I_2(m;T^\prime)$ for three different regimes depending on the relationship between the maximum $m$ to some parameter $T^\prime$, see Eq.~(\ref{secondcondition}). The most relevant integration is (a). Our claim is that in this case we may restrict the integration in Eq.~(\ref{whatisintegral}) to $m=T^\prime$, since the constraint limits the relevant domain of the integration variables.}\label{fig:two}
\end{figure}

In order to finish the proof of Eq.~(\ref{convtheorem}) we show it for $N+1$ while assuming that the statement is true for $N$. We write again the definition of the integral 
\begin{equation}\begin{split}
&I_{N+1}(m,T^\prime) \\
&= \int\limits_0^m d\tau_1 \ldots \int\limits_0^m d\tau_{N+1}  \prod_{i=1}^{N+1} g_i(\tau_i)\delta\left(T^\prime - \sum_{j=1}^{N+1} \tau_j\right).
\end{split}\end{equation}
We rearrange the order of integration and separate $-\tau_N+1$ in the delta function 
\begin{equation}\begin{split}\label{gettt}
&I_{N+1}(m,T^\prime)\\
& = \int\limits_0^m d\tau_{N+1}g_{N+1}(\tau_{N+1})\\
&\times\left[ \int\limits_0^m d\tau_1 \ldots \int\limits_0^m d\tau_N  \prod_{i=1}^N g_i(\tau_i)\delta\left(T^\prime -\tau_{N+1} - \sum_{j=1}^N \tau_j\right) \right] 
\end{split}\end{equation}
Now we consider $T^\prime<m$ which corresponds to regime (a) from Eq.~(\ref{secondcondition}). From this it is obviously $T^\prime-\tau_{N+1}<m$ because $T^\prime-\tau_{N+1}<T^\prime$. This inequality $T^\prime-\tau_{N+1}<m$ is exactly the condition for which the $N$-multiple integral inside the square bracket of Eq.~(\ref{gettt}) is the $N$-fold convolution
\begin{equation}\begin{split}\label{stepstep}
I_{N+1}(m,T^\prime) &= \int\limits_0^m d\tau_{N+1} g_{N+1}(\tau_{N+1})\\
&\times\left[(g_1\ast\ldots\ast g_N)^{(N)} (T^\prime-\tau_{N+1}) \right] 
\end{split}\end{equation}
according to the assumption of the induction proof. The remaining integral over $\tau_{N+1}$ is zero from $T^\prime$ to $m$. The difference $T^\prime-\tau_{N+1}=\sum_{i=1}^N \tau_i$ is positive because all $\tau_i$ are positive. So when $\tau_{N+1}>T^\prime$ we cannot fulfill the constrain. This property is controlled by the convolution in the integrand of Eq.~(\ref{stepstep}) which is zero for negative arguments. So we get
\begin{equation}\begin{split}
I_{N+1}(m,T^\prime) &= \int\limits_0^{T^\prime} g_{N+1}(\tau_{N+1})\\
&\times d\tau_{N+1} \left[(g_1\ast\ldots\ast g_N)^{(N)} (T^\prime-y_{N+1}) \right] 
\end{split}\end{equation}
and this is the convolution. Remember that we assumed $T^\prime<m$ in Eq.~(\ref{stepstep}). Therefore we showed Eq.~(\ref{convtheorem}).\\

 With the same arguments Eq.~(\ref{convtheorem}) can also be stated for discrete random variables with some arbitrary constraint $C^\prime>0$. It is equivalently
\begin{equation}\label{convtheorem2}
\sum_{y_1=0}^m \ldots \sum_{y_N=0}^m \prod_{i=1}^N g_i(y_i)\delta_{C^\prime,\sum_{j=1}^N y_j} =  (g_1\ast\ldots\ast g_N)^{(N)}(C^\prime)
\end{equation}
for $m>C^\prime$. For \textit{ZRP} it is $y_i=\kappa_i$ and $C^\prime=K-m$ and for \textit{TIDSI} it is $y_i=\lambda_i$ and $C^\prime=L-m$. For \textit{ZRP} and \textit{TIDSI} the functions are $g_i=\psi$ for all $i$.

\section{Typical fluctuations of \textit{TIDSI} for the parameter $\alpha\in(0,1)$}\label{sec:appb}
In \cite{bar2016exact} the typical fluctuations of $f(m;L)$ in the second half $L/2<m<L$ where calculated as
\begin{equation}\label{bar}
L f(m;L) \sim \frac{1}{\xi^2} \frac{d}{du}H(u)|_{u=1/\xi}
\end{equation}
with the function
\begin{equation}\begin{split}
H(u)&= \frac{\Gamma(\alpha)}{\Gamma(2\alpha+1)|\Gamma(-\alpha)|}u^{1-\alpha} (u-1)^{2\alpha}\\
&\times {}_2F_1(1,1+\alpha,1+2\alpha,1-u).
\end{split}\end{equation}
The hypergeometric function defined as
\begin{equation}
{}_2F_1(a,b,c,z) = \sum_{j=0}^\infty \frac{(a)_j(b)_j}{(c)_j} \frac{z^j}{j!}
\end{equation}
with the Pochhammer symbol $(a)_j=\Gamma(a+j)/\Gamma(a)$. \\

We show now that Eq.~(\ref{bar}) is identical to our result from Eq.~(\ref{our}). For that let us first take the derivative of the right hand side of Eq.~(\ref{bar}) while $u=1/\xi$:
\begin{equation}\begin{split}
u^2 \frac{d}{du}H(u) &= u^2 \frac{\Gamma(\alpha)}{\Gamma(2\alpha+1)|\Gamma(-\alpha)|}\\
&\Big[\left[(1-\alpha)u^{-\alpha}(u-1)^{2\alpha} +2\alpha u^{1-\alpha}(u-1)^{2\alpha-1} \right]\\
&\times {}_2F_1(1,1+\alpha,2\alpha+1,1-u) \\
&-\frac{1+\alpha}{1+2\alpha}u^{1-\alpha}(u-1)^{2\alpha}\\
&\times {}_2F_1(2,2+\alpha,2\alpha+2,1-u)\Big]
\end{split}\end{equation}
where we used $d/dz{}_2F_1(a,b,c,z) = ab/c {}_2F_1(1+a,1+b,1+c,z)$. Now we take out the term $u^{-\alpha}(u-1)^{2\alpha-1}$ so that
\begin{equation}\begin{split}\label{sidestep}
u^2 \frac{d}{du}H(u)& = \frac{\Gamma(\alpha)}{\Gamma(2\alpha+1)|\Gamma(-\alpha)|}u^{2-\alpha}(u-1)^{2\alpha-1}\\
&\Big[\left[(1-\alpha)(u-1) +2\alpha u \right]\\
&\times {}_2F_1(1,1+\alpha,2\alpha+1,1-u) \\
&-\frac{1+\alpha}{1+2\alpha}u(u-1)\\
&\times{}_2F_1(2,2+\alpha,2\alpha+2,1-u)\Big].
\end{split}\end{equation}
To show the identity to Eq.~(\ref{our}) we have to show that the expression inside the big squared bracket of Eq.~(\ref{sidestep}) is identical to $2\alpha$. Let us write this question shortly as
\begin{equation}\label{shortly}
f(u)F(1,1-u)+g(u)F(2,1-u) = 2\alpha,
\end{equation}
i.e. is this statement true?
Here $f(u)=(1-\alpha)(u-1) +2\alpha u $, $g(u)=-(1+\alpha)/(1+2\alpha)u(u-1)$ and $F(i,1-u)={}_2F_1(i+1,i+1+\alpha,i+1+2\alpha,1-u)$.\\
 
Since the hypergeometric function depends on $1-u$ we consider the series expansion at $u=1$ of the inner bracket. In principle any other point could be considered but the problem becomes simpler at $u=1$. The Taylor series of Eq.~(\ref{shortly}) is
\begin{equation}\begin{split}
&f(u)F(1,1-u)+g(u)F(2,1-u) \\
&=\sum_{j=0}^\infty \Bigg( f(u)F(1,1-u)+g(u)F(2,1-u)\Bigg)^{(j)}\Big|_{u=1} \\
&\times \frac{(u-1)^j}{j!}.
\end{split}\end{equation}
We apply the general Leibniz rule of derivation
\begin{equation}\begin{split}\label{leibniz}
&f(u)F(1,1-u)+g(u)F(2,1-u)  \\
 &=\sum_{j=0}^\infty\Bigg( \sum_{k_1=0}^j {j\choose{k_1}} F^{(j-k_1)}(1,1-u) f^{(k_1)}(u)\Big|_{u=1}\\
&+\sum_{k_2=0}^j {j\choose{k_2}} F^{(j-k_2)}(2,1-u) g^{(k_2)}(u)\Big|_{u=1}\Bigg) \frac{(u-1)^j}{j!}.
\end{split}\end{equation}
The derivatives of $f$ and $g$ are 
\begin{equation}\begin{split}\label{derivfg}
f^{(k_1)}(u)\Big|_{u=1} &= 
\begin{cases}
2\alpha & \text{ for } k_1=0,\\
1+\alpha & \text{ for } k_1=1,\\
0 & \text{ for } k_1\ge 2,
\end{cases}\\
g^{(k_2)}(u)\Big|_{u=1} &= 
\begin{cases}
0 & \text{ for } k_2=0,\\
-\frac{1+\alpha}{1+2\alpha} & \text{ for } k_2=1,\\
-2\frac{1+\alpha}{1+2\alpha}& \text{ for } k_2= 2,\\
0 & \text{ for } k_2\ge 3.
\end{cases}
\end{split}\end{equation}
The two sums in Eq.~(\ref{leibniz}) are only nonzero for $k_1=0,1$ and $k_2=2,3$. Thus we can write
\begin{equation}\begin{split}
&f(u)F(1,1-u)+g(u)F(2,1-u)\\
& =\sum_{j=0}^\infty \Bigg( \sum_{k_1=0}^1 {j\choose{k_1}} F^{(j-k_1)}(1,1-u) f^{(k_1)}(u)\Big|_{u=1}\\
&+\sum_{k_2=1}^2 {j\choose{k_2}} F^{(j-k_2)}(2,1-u) g^{(k_2)}(u)\Big|_{u=1}\Bigg) \frac{(u-1)^j}{j!}.
\end{split}\end{equation}
The binomial is zero when $k_1>j$ and $k_2>j$ so this expression is valid for all $j$. Now we express the hypergeometric function $F(2,1-u)$ by $F(1,1-u)$ via the relationship of their derivatives. The $j$-th derivative of the hypergeometric function at $u=1$ is
\begin{equation}
F^{(j)}(1,1-u)|_{u=1} = (-1)^j \frac{(1)_j(1+\alpha)_j}{(1+2\alpha)_j},
\end{equation}
thus
\begin{equation}
F^{(j)}(2,1-u)\Big|_{u=1} = - \frac{1+2\alpha}{1+\alpha}F^{(j+1)}(1,1-u).
\end{equation}
So we can write
\begin{equation}\begin{split}
&f(u)F(1,1-u)+g(u)F(2,1-u) \\
&=\sum_{j=0}^\infty \Bigg( \sum_{k_1=0}^1 {j\choose{k_1}} F^{(j-k_1)}(1,1-u) f^{(k_1)}(u)\Big|_{u=1}\\
&-\frac{1+2\alpha}{1+\alpha}\\
&\times\sum_{k_2=1}^2 {j\choose{k_2}} F^{(j-k_2+1)}(1,1-u) g^{(k_2)}(u)\Big|_{u=1}\Bigg) \frac{(u-1)^j}{j!}.
\end{split}\end{equation}
We order according to the hypergeometric functions
\begin{equation}\begin{split}
&f(u)F(1,1-u)+g(u)F(2,1-u)\\
& =\sum_{j=0}^\infty \Bigg( F^{(j)}(1,1-u)\\
&\times \Bigg[ {j\choose{0}} f^{(0)}(u)-\frac{1+2\alpha}{1+\alpha} {j\choose{1}} g^{(1)}(u) \Bigg] \Big|_{u=1}\\
&+F^{(j-1)}(1,1-u) \\
&\times\Bigg[ {j\choose{1}} f^{(1)}(u)-\frac{1+2\alpha}{1+\alpha} {j\choose{2}} g^{(2)}(u) \Bigg] \Big|_{u=1} \Bigg) \frac{(u-1)^j}{j!}.
\end{split}\end{equation}
With Eq.~(\ref{derivfg}) we get
\begin{equation}\begin{split}\label{nextcalc}
&f(u)F(1,1-u)+g(u)F(2,1-u) \\
&=\sum_{j=0}^\infty \Bigg( (2\alpha+j)F^{(j)}(1,1-u)  \Big|_{u=1}\\
&+j(\alpha+j)F^{(j-1)}(1,1-u)\Big|_{u=1} \Bigg) \frac{(u-1)^j}{j!}.
\end{split}\end{equation}
Now we split the summation over $j$ for $j=0$ and all other $j\ge 1$. For the latter we use the relationship between successive orders of the derivative for the hypergeometric function
\begin{equation}\label{relder}
F^{(j)}(1,1-u)\Big|_{u=1} = - \frac{j(\alpha+j)}{2\alpha+j} F^{(j-1)}(1,1-u)\Big|_{u=1}
\end{equation}
valid for $j\ge 1$. This gives zero for all terms with $j\ge 1$ in Eq.~(\ref{nextcalc}) and only the term with $j=0$ remains. With $F^{(0)}(1,1-u)|_{u=1}=1$ we obtain
\begin{equation}
f(u)F(1,1-u)+g(u)F(2,1-u)=2\alpha 
\end{equation}
Thus we finally showed that indeed
\begin{equation}
u^2 \frac{d}{du} H(u) = \frac{\Gamma(\alpha)}{\Gamma(2\alpha)|\Gamma(-\alpha)|} u^{2-\alpha}(u-1)^{2\alpha-1}.
\end{equation}
Hence Eq.~(\ref{bar}) is identical to our result from Eq.~(\ref{our}).

\end{appendix}

\bibliographystyle{prestyle}
\bibliography{citations2}

\end{document}